\documentclass[aps,prd,reprint,twocolumn,preprintnumbers,floatfix,nofootinbib]{revtex4-1}
\usepackage[utf8]{inputenc}
\usepackage[dvips]{graphicx}
\usepackage[dvipsnames]{xcolor}
\usepackage{color}
\usepackage{relsize}
\usepackage{graphics}
\usepackage{epstopdf}
\usepackage{hyperref}
\usepackage{mathrsfs}
\usepackage{amssymb}

\usepackage[normalem]{ ulem }
\usepackage{amsthm}
\usepackage{amsmath}
\usepackage{cancel}
\usepackage[caption=false]{subfig}
\usepackage{tabularx,ragged2e} 
 \newcolumntype{L}{>{\RaggedRight\arraybackslash}X}

\usepackage{hyperref}

\usepackage[normalem]{ ulem }
\usepackage{amsthm}
\usepackage{amsmath}
\usepackage{hyperref}
\usepackage{cancel}

\newcommand{\be}{\begin{equation}}
\newcommand{\ee}{\end{equation}}
\newcommand{\bea}{\begin{eqnarray}}
\newcommand{\eea}{\end{eqnarray}}

\usepackage[compat=1.0.0]{tikz-feynman}

\begin{document}
\sloppy

\preprint{IPPP/23/17}

\vspace*{1mm}

\title{Hunting for Neutral Leptons with Ultra-High-Energy Cosmic Rays}

\author{Robert Heighton$^{a}$}
\email{robert.heighton@durham.ac.uk}
\author{Lucien Heurtier$^{a}$}
\email{lucien.heurtier@durham.ac.uk}
\author{Michael Spannowsky$^{a}$}
\email{michael.spannowsky@durham.ac.uk}

\affiliation{\footnotesize $^a$ Institute for Particle Physics Phenomenology, Durham University, South Road, Durham, U.K.}

\begin{abstract}
Next-generation large-volume detectors, such as  GRAND, POEMMA, Trinity, TAROGE-M, and PUEO, have been designed to search for ultra-high-energy cosmic rays (UHECRs) with unprecedented sensitivity. 
We propose to use these detectors to search for new physics beyond the Standard Model (BSM). By considering the simple case of a right-handed neutrino that mixes exclusively with the active $\tau$ neutrino, we demonstrate that the existence of new physics can increase the probability for UHECRs to propagate through the Earth and produce extensive air showers that will be measurable soon. We compare the fluxes of such showers that would arise from various diffuse and transient sources of high-energy neutrinos, both in the Standard Model and in the presence of a right-handed neutrino. We show that detecting events with emergence angles $\gtrsim 10$ deg is promising to probe the existence of BSM physics, and we study the sensitivity of GRAND and POEMMA to do so. In particular, we show that the hypothesis of a right-handed neutrino with a mass of $\mathcal O(1-16)$ GeV may be probed in the future for mixing angles as small as $|U_{\tau N}|^2 \gtrsim 10^{-7}$, thus competing with existing and projected experimental limits.
\end{abstract}
\maketitle

\section{Introduction}\label{sec:intro}

The search for ultra-high-energy neutrino cosmic rays (UHECRs) has attracted more and more attention in the last decades as new generations of ground-based detectors such as the Auger observatory, IceCube, and ANTARES and balloon experiments such as ANITA have begun to use volumes large enough to probe the existence of the so-called Greisen–Zatsepin–Kuzmin (GZK) spectrum. This diffuse flux of ultra-high-energy (UHE) neutrinos, originating from the scattering of proton cosmic rays on the cosmic microwave background, is predicted to spread over energies as large as $\mathcal O(10-100)$ exaelectronvolts (EeV). Although the GZK spectrum remains undetected, the increasing sensitivity of future large-volume detectors such as POEMMA~\cite{Venters:2019xwi, POEMMA:2020ykm}, GRAND~\cite{Lago:2021dom}, Trinity~\cite{Wang:2021zkm}, TAMBO~ \cite{Zhelnin:2022ybr}, or TAROGE-M~\cite{TAROGE:2022soh} is likely to render its detection possible within the next few decades. 

Transient sources, which can last from hundreds of seconds to a few months, constitute another compelling source of UHE neutrinos in the cosmos. By injecting significant energy over a relatively short time, they constitute some of the most promising sources with sufficient energy and flux to be detectable. Experimentally speaking, the search for a time-dependent neutrino source is also known to reduce the background due to atmospheric neutrinos and muons~\cite{IceCube:2015usw, ANTARES:2015gxt}, favouring their detection compared to diffuse fluxes. Active galactic nuclei, neutron-star/black-hole mergers, and gamma-ray bursts (GRBs) are examples of such transients (see Ref.~\cite{Guepin:2022qpl} and references therein for a recent review) that will become increasingly important in the years to come, as they constitute crucial channels for discovery in the field of multi-messenger astronomy. 

For a transient event that would occur at a distance of $\mathcal O(10)$ Mpc, future detectors predict the observation of a large number of events. Indeed, GRAND, POEMMA, Trinity, TAROGE-M, and PUEO may each see hundreds of UHE neutrinos in such a case~\cite{Venters:2019xwi, PUEO:2020bnn, Wang:2021zkm, TAROGE:2022soh, POEMMA:2020ykm, Guepin:2022qpl}. Depending on the location of this transient source in the Universe and the location of the detector on (or around) the Earth, this number of events may vary, as the chord length between the point of incidence of a UHECR entering the Earth and the point where it reaches the detector may vary. In the Standard Model (SM) of particle physics, UHE neutrinos may only propagate through the Earth for relatively short distances ($\lesssim \mathcal O(100)$ km). Indeed, the neutrino-nucleon scattering cross-section increases with the energy of the incoming ray, and its mean-free path through the Earth decreases accordingly. For this reason, the detection of the GZK spectrum, or any transient source that is close enough from the Earth, is most likely to be detected first via the observation of UHECRs that do not traverse long distances through the Earth before hitting a detector, but instead enter the Earth's surface with small incidence angles. 

Sensors such as GRAND and Trinity were designed primarily to measure such Earth-skimming neutrino UHECRs with the highest possible accuracy. However,  other detectors, such as the existing collaboration ANITA, its upgrade PUEO, or the space-based project POEMMA, typically observe upward-propagating cosmic rays from a very high altitude and can therefore observe UHECRs that could, in principle, exit the Earth with larger incidence/emergence angles. A few years ago, the ANITA collaboration detected events that were interpreted as upward propagating but featured large emergence angles ($\gtrsim 30^{\circ}$) and were thus reported as anomalous, as they were not associated with any point-like transient emission. This claim triggered much attention from particle physics theorists, since SM neutrinos are unlikely to propagate over such large distances through the Earth~\cite{Romero-Wolf:2018zxt}. Instead, new physics beyond the Standard Model (BSM) may lead to different predictions, which led theorists to interpret these ANITA events as a hint of new physics~\cite{Heurtier:2019git, Heurtier:2019rkz, Anchordoqui:2018ucj, Huang:2018als, Chauhan:2018lnq, Cherry:2018rxj, Collins:2018jpg, BhupalDev:2020zcy, Borah:2019ciw, Cline:2019snp, Anchordoqui:2018ssd, Esmaili:2019pcy, Esteban:2019hcm}.

In this work, we demonstrate, using the simple case of a $\mathcal O(\mathrm{GeV})$ right-handed neutrino, that detecting neutrino UHECRs with large incidence/emergence angles in the context of multi-messenger astronomy can serve as a compelling test of new physics scenarios. In particular, several studies have pointed out that new physics could affect the propagation and energy loss of UHECRs when propagating through the Earth~\cite{Heurtier:2019git, Heurtier:2019rkz}. It could alter the angular distribution of the extensive air showers (EAS) measured by ANITA, for instance, as compared to SM predictions. This idea was explored by considering the possible modification of the neutrino-nuclei cross-section due to the presence of new physics~\cite{GarciaSoto:2022vlw, Denton:2020jft, Huang:2021mki}. The authors of Ref.~\cite{Denton:2020jft} showed that measuring the angular distribution of a hundred events by GRAND or POEMMA could help to measure this cross-section with a 20\% accuracy. Here we also envision that the propagation of new physics states may help UHECRs to make it through the Earth with long chord lengths, and study how this can be used to prove the existence of new physics using large-volume detectors.

The paper is organised as follows: In Sec.~\ref{sec:model}, we introduce the particle physics model we consider throughout this work. We then describe in Sec.~\ref{sec:CRpropagation} how the presence of a right-handed neutrino in the theory affects the propagation of CRs through the Earth. In Sec.~\ref{sec:detector} we discuss which detectors are most sensitive to such effects and compute their corresponding effective areas. Finally, we define in Sec.~\ref{sec:strategy} our searching strategy and present our results in Sec.~\ref{sec:results}, before concluding in Sec.~\ref{sec:conclu}.

\section{The Model}\label{sec:model}
The minimal extension of the SM leptonic sector we consider contains the three generations of SM left-handed $SU(2)_L$ doublets $L_\alpha$ ($\alpha=e,\mu,\tau$), to which we add a right-handed SM singlet, assumed to be a Majorana fermion and denoted by $N$. Throughout this work, we consider the simple case where this right-handed neutrino mixes exclusively with the active $\tau$ neutrino. This corresponds to simply adding the following contribution to the SM lagrangian:
\bea
  -\mathcal{L} &\supset& -\frac{m_N}{2}\bar N^c N+ \frac{g}{\sqrt{2}} \sin{\theta_{\text{mix}}} W_{\mu}^{+} \bar{N}^{c} \gamma^{\mu} P_{L} \tau \nonumber\\
  &&+ \frac{g}{2\cos{\theta_w}} \sin{\theta_{\text{mix}}} Z_{\mu} \bar{N}^{c} \gamma^{\mu} P_{L} \nu_{\tau} + \text{h.c.}
\eea
where $m_N$ and $\theta_{\text{mix}}$ denote the mass of the right-handed neutrino and its mixing angle with active neutrinos, respectively. 

\begin{figure}
    \centering
    \includegraphics[width=\linewidth]{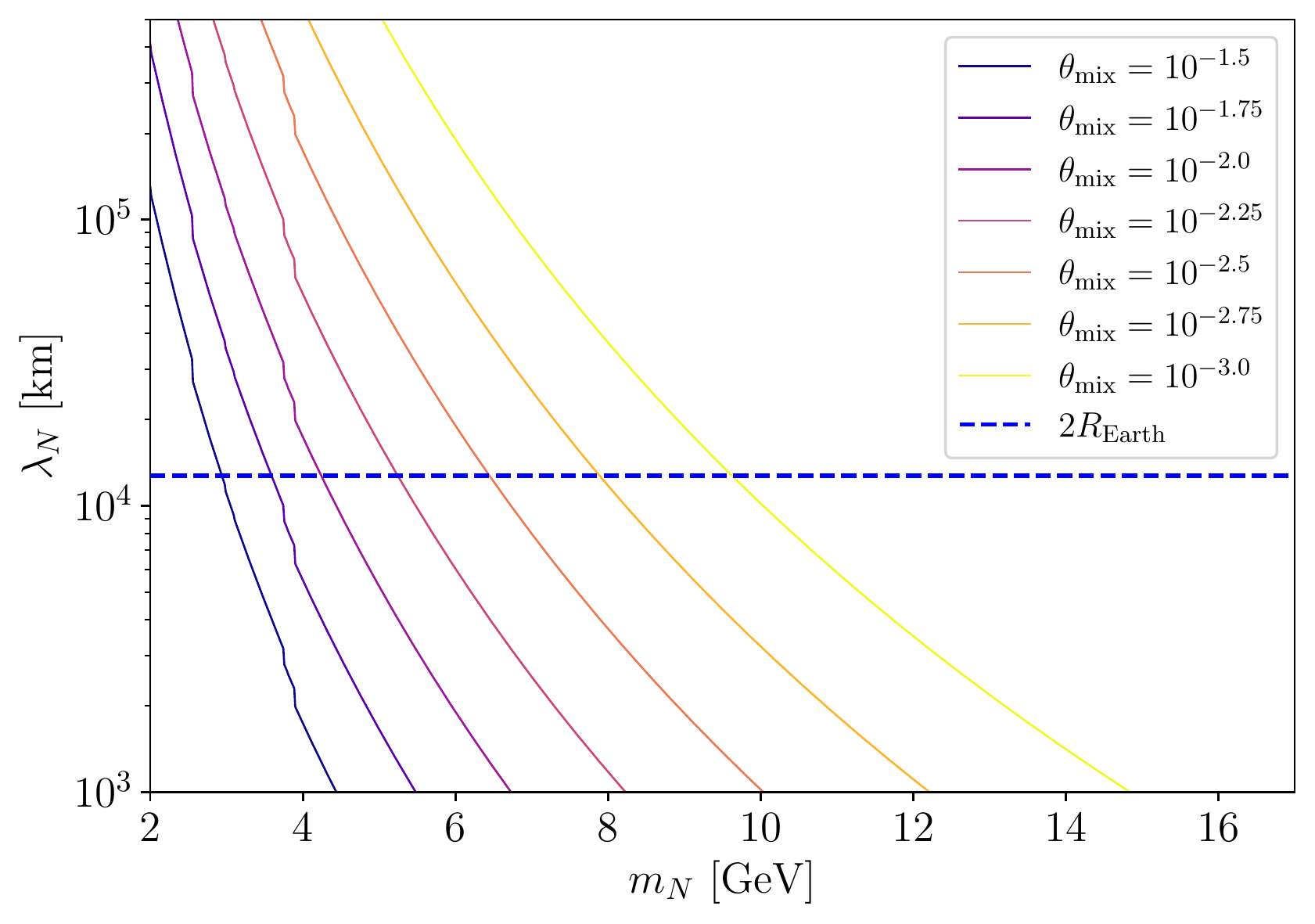}
    \caption{\label{fig:decay_length} \footnotesize The decay length of a right-handed neutrino of energy $E$ = 10 EeV as a function of its mass $m_N$ in a model with mixing angle $\theta_{\text{mix}}$. The diameter of the Earth, $2R_{\text{Earth}}$, is provided for reference.}
\end{figure}

\vspace{5pt}
\noindent{\bf Decay Modes.} Depending on its mass, a heavy RH neutrino has access to various 2 and 3-body decay channels (see e.g. App C of Ref.~\cite{Atre:2009rg}). At ultra-high energy $E\gg m_N$, the right-handed neutrino is highly boosted, and its corresponding decay length scales linearly with $E$ and quadratically with the inverse of its mixing angle $\theta_{\rm mix}$, and can be parametrized as
\begin{equation}
\lambda_N (E) \approx \left( \frac{E}{\text{EeV}} \right) \left( \frac{\theta_{\text{mix}}}{0.01} \right)^{-2} L(m_N)\,,
\end{equation}
where $L(m_N)$ corresponds to its decay length calculated at \mbox{$E=1$ EeV} and for $\theta_{\rm mix}=10^{-2}$. In Fig.~\ref{fig:decay_length}, we represent the evolution of $\lambda_N(E)$ as a function of the mass $m_N$ and mixing angle $\theta_{\rm mix}$. In the figure, we compare the value of the Earth's diameter. One may note that for the range of mixing angles 
 and the incoming energy considered, RHNs whose decay length is comparable to this value may help UHECRs propagate over sizeable distances through the Earth, have masses of $\mathcal O(1-10)$ GeV. 
 
\begin{figure*}    \includegraphics[width=\linewidth]{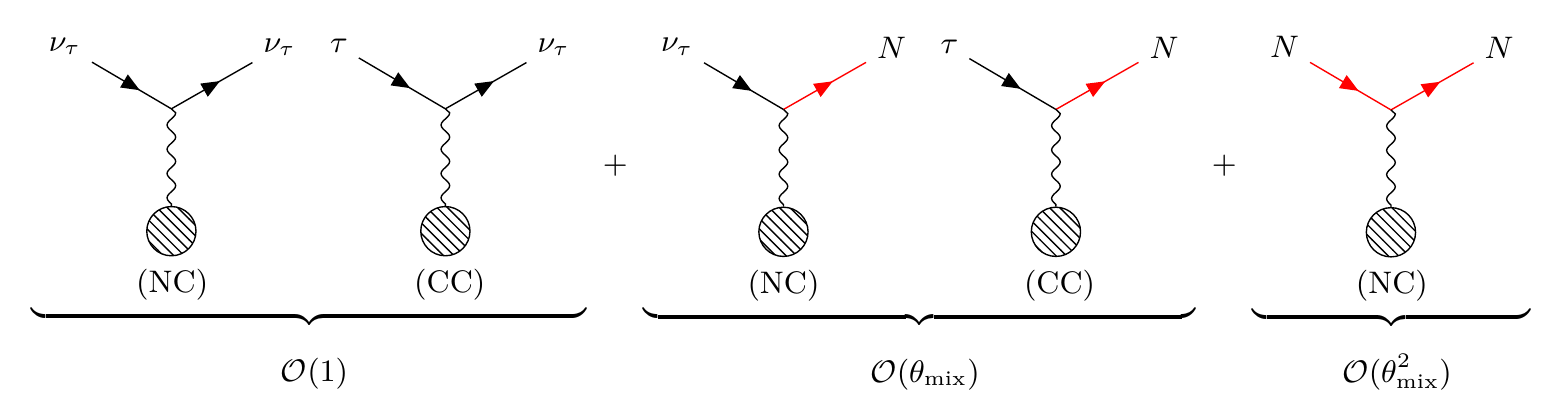}
    \caption{\label{fig:diagram_interactions} \footnotesize Diagrams involving $\tau$ leptons, $\tau$-neutrinos, and the RH neutrino $N$ in the scattering off a nucleon at rest.}
\end{figure*}

~\newline
\noindent{\bf Scattering.} Given the simplicity of the model, the only way the RH neutrino can scatter off ordinary matter is through neutral current (NC) and charged current (CC) interactions that are inherited from the SM neutrino interactions. As can be seen from Fig.~\ref{fig:diagram_interactions}, such interactions scale like $(\sin\theta_{\rm mix})^n$ for diagrams involving $n$ RH neutrinos.
In practice, the energy of the ultra-high-energy cosmic rays that we consider throughout this work is so large that leptons scattering off a nucleon are sensitive to its internal structure, corresponding to a deep inelastic scattering. Given the corresponding cross sections $\sigma_{\rm NC}$ and $\sigma_{\rm CC}$ (for NC and CC interactions of neutrinos) as calculated in the SM, one can obtain the values of the cross sections $\sigma_{\text{NC,mix}}$ and $\sigma_{\text{CC,mix}}$ involving RH neutrinos using the appropriate scaling with the mixing angle, as follows:
\begin{equation}\label{eq:cross_sections}
\begin{aligned}
\\
\sigma_{\text{NC,mix}} &= \sigma_{\text{NC}} \sin^{2}(\theta_{\text{mix}})\,, \\
\sigma_{\text{CC,mix}} &= \sigma_{\text{CC}} \sin^{2}(\theta_{\text{mix}})\,. \\
\\
\end{aligned}
\end{equation}

\section{UHE Neutrino Propagation}\label{sec:CRpropagation}

\subsection{The SM Case}

In the Standard Model case, UHE neutrinos incident on the Earth are likely to interact with nucleons, and may do so via NC or CC interactions, corresponding to the $\mathcal{O}(1)$ diagrams in Fig.~\ref{fig:diagram_interactions}. In the NC case, a neutrino will lose energy but remain a neutrino; in the CC case, the neutrino will both lose energy and convert to a charged lepton. In this work, we primarily investigate $\tau$ neutrinos, $\nu_{\tau}$, and so the particle produced in a CC interaction is a $\tau$ lepton.

The neutrino-nucleon scatterings cross sections of these processes  scale with the neutrino energy $E_{\nu}$ according to the power-law approximations
\begin{equation}\label{eq:energy_scaling}
\begin{aligned}
\\
\sigma_{\text{NC}} &\approx (2.31 \times 10^{-36} \text{cm}^{2}) \left( \frac{E_{\nu}}{\text{GeV}} \right)^{\alpha}  \\
\sigma_{\text{CC}} &\approx (5.53 \times 10^{-36} \text{cm}^{2}) \left( \frac{E_{\nu}}{\text{GeV}} \right)^{\alpha} \\
\\
\end{aligned}
\end{equation}
where the index $\alpha \approx 0.363$ \cite{Formaggio:2012}, and so the Earth becomes more opaque to the neutrino flux at higher energies. Neutrino-electron interactions are subdominant at such energies \cite{Formaggio:2012} and are neglected here.

The neutrino resulting from an NC interaction will likely undergo further scatterings (NC and CC) as it traverses a chord of the Earth's interior. The charged $\tau$ lepton produced by a CC interaction will continue to propagate through the Earth, undergoing various energy losses via effects such as bremsstrahlung and photonuclear interactions \cite{Koehne:2013}, and will eventually decay, producing a $\tau$ neutrino in a process known as $\nu_\tau$ {\em regeneration}. The additional decay products comprise a pair of leptons (one charged and one neutrino) of lighter flavour; these are neglected, as neutrinos and charged leptons of lighter flavours cannot regenerate $\tau$ neutrinos through their interactions and decays.

As mentioned, an emergent effect arising from the combination of CC interactions and $\tau$ decays is $\nu_\tau$ regeneration, whereby a $\tau$ neutrino initially scatters with a nucleon and converts to a $\tau$ via a CC interaction, and later, after propagating some distance, subsequently decays to revert to a $\tau$ neutrino of lesser energy than that with which it began. The process of $\nu_\tau$ regeneration significantly impacts the flux and energy spectrum of Earth-traversing UHE neutrinos \cite{Alvarez_Muniz:2018}.

The general principle of a regenerative effect (due to temporary propagation as a different particle species) motivates this work. Indeed, in the SM, $\nu_\tau$ regeneration is the dominant source of EAS above the EeV scale at emergence angles $\gtrsim \mathcal O(10^\circ)$ (see e.g.~\cite{Alvarez_Muniz:2018}). Under the hypothesis of a BSM scenario involving particles long-lived enough to propagate over distances comparable to the $\tau$ decay length at such energies, it is thus reasonable to believe that BSM physics could play an important role in the propagation of UHECRs at such large emergence angles. This is the nature of our investigation in the case of our minimal right-handed neutrino model, as the RHN may play the role of such a particle, providing an intermediate through which an Earth-traversing UHECR can significantly regenerate.
Large-volume detectors would then constitute essential observatories for searching for the existence of potential new physics.

\begin{figure*}[t!]
    \includegraphics[width=0.49\linewidth]{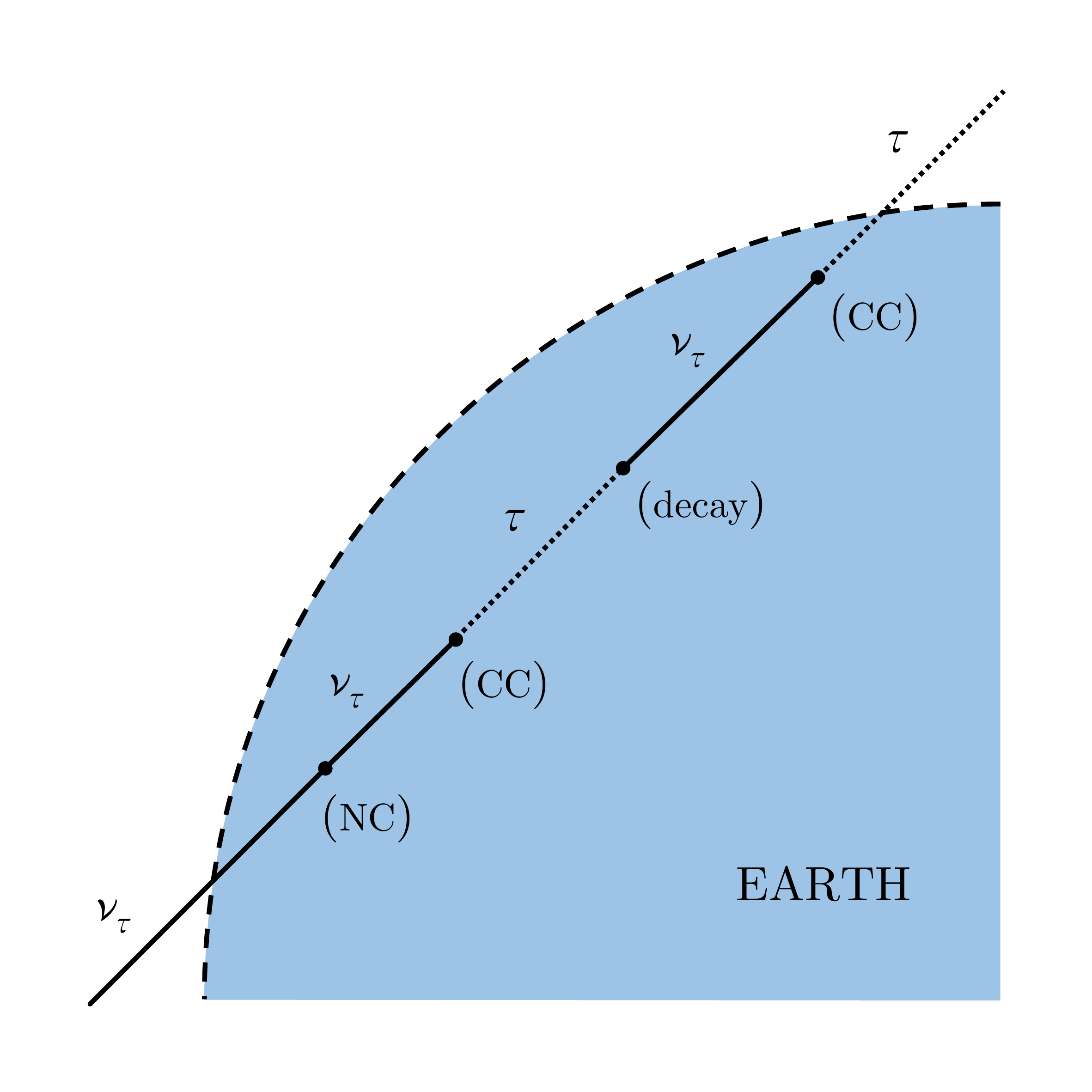}\includegraphics[width=0.49\linewidth]{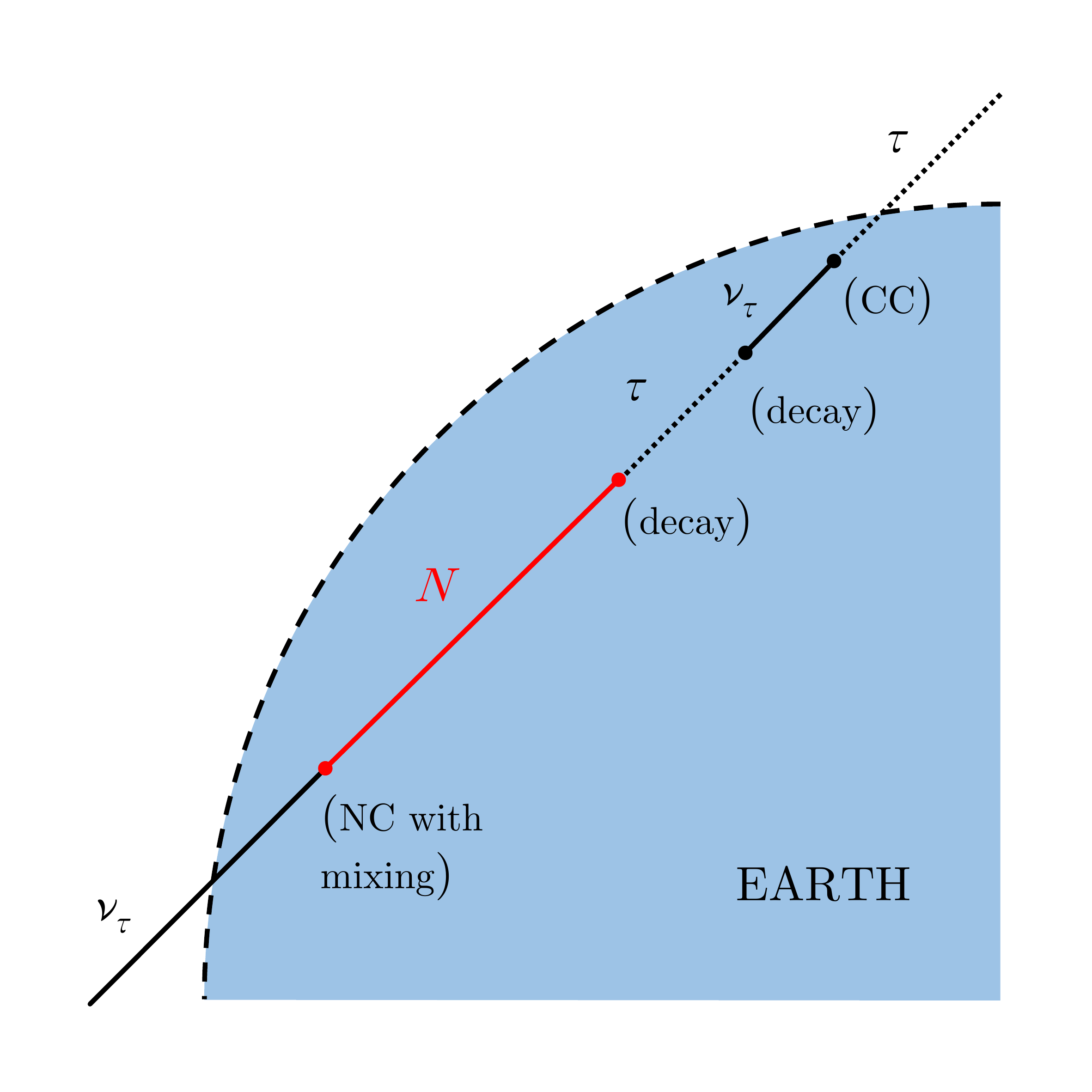}
    \caption{\label{fig:earthdiagram} \footnotesize Example tracks for an Earth-traversing UHE $\tau$ neutrino in the context of the Standard Model (left panel) and under the influence of our BSM model (right panel). Black solid lines and dotted lines depict the particle propagating as a $\tau$ neutrino and charged lepton  respectively, with larger black dots representing interactions and decays as labelled. The temporary propagation of the particle as a $\tau$ before re-converting to a $\nu_{\tau}$ constitutes an example of $\nu_\tau$ regeneration. On the right panel, the propagation of the right-handed neutrino $N$ is represented in red. The temporary propagation of the particle as a RHN before reverting to a $\tau$ or $\nu_{\tau}$ effectively contributes to the $\nu_\tau$ regeneration.}  
\end{figure*}

\subsection{Adding the RHN}

We now consider the propagation of ultra-high-energy neutrinos in the Earth under the influence of the BSM model described in Sec.~\ref{sec:model}.

In addition to the usual NC and CC interaction possibilities, a left-handed neutrino traversing the Earth may now additionally undergo a \textit{mixing} NC interaction, corresponding to the first $\mathcal{O}(\theta_{\text{mix}})$ diagram in Fig.~\ref{fig:diagram_interactions}. This is largely analogous to the Standard Model equivalent, excepting that where the SM case had the particle remain a left-handed neutrino, $\nu_{\tau}$, the outgoing particle of the mixing interaction is the right-handed neutrino $N$. Relative to that of the NC interaction, the cross-section of the mixing NC interaction is suppressed by a factor of $\sin^{2}{\theta_{\text{mix}}}$ as expressed in Eq.~\eqref{eq:cross_sections}.

The right-handed neutrino produced by such a scattering continues propagating through the Earth. Like its left-handed counterpart, it may scatter with a nucleon via two possible interactions: a mixing NC interaction whereby it reverts to a $\nu_{\tau}$, and a mixing CC interaction whereby it becomes a charged $\tau$ lepton (which propagates and ultimately decays back to a $\nu_{\tau}$ as in the SM case). The cross sections of both interactions are similarly suppressed by a factor of $\sin^{2}{\theta_{\text{mix}}}$, corresponding to the two $\mathcal{O}(\theta_{\text{mix}})$ diagrams in Fig.~\ref{fig:diagram_interactions}.

In principle, $N$ may undergo a doubly-mixing NC interaction whereby it remains a right-handed neutrino, corresponding to the $\mathcal{O}(\theta_{\text{mix}}^2)$ diagram in Fig.~\ref{fig:diagram_interactions}, but the cross-section of such scattering is suppressed by a factor of $\sin^{4}{\theta_{\text{mix}}}$, and is therefore neglected in our simulation on statistical grounds.

Also available to the right-handed neutrino, and generally dominating over the scatterings in our findings, is RHN decay, as briefly discussed in Sec.~\ref{sec:model}. The decay width is strongly dependent on the RHN mass $m_N$, with terms in ${m_{\text{N}}}^3$ and ${m_{\text{N}}}^5$, and with larger masses opening up numerous new hadronic decay channels. RHN decay produces $\tau$ neutrinos and charged $\tau$ leptons that propagate onward within the Earth, behaving as before. A comparison may be drawn between the production and subsequent decay of an RHN and the SM process of $\tau$ regeneration.

In the SM case, a single initial $\nu_{\tau}$ incident on the Earth may be regarded (and hence simulated) consistently as a single particle, undergoing interactions and converting between $\nu_{\tau}$ and $\tau$ (while daughter products of lighter leptonic flavours are neglected), but remaining one particle `instance' and resulting in at most one $\nu_{\tau}$ or $\tau$ exiting the Earth. In the RHN case, however, some RHN decay channels produce multiple instances of the relevant particles, including, for example,

\begin{equation}
    \begin{aligned}
    N \rightarrow \nu_{\tau} \tau^+ \tau^- , \\
    \end{aligned}
\end{equation}

and so in the context of the BSM model, it is necessary to account for the proliferation of a single particle instance into multiple; the production and subsequent decay of a right-handed neutrino may result in two or more detectable particles stemming from the same initial $\nu_{\tau}$.

The decay of the RHN also constitutes a new avenue for detection. Analogously with the $\tau$, atmospheric decay via hadronic channels is likely to instigate an extensive air shower (EAS) that may be observed by detectors such as POEMMA and GRAND. Hence, in simulating the propagation described in this Section and computing effective areas and results for detectors, we must consider both the usual $\tau$ and the potential observable decay of the RHN.

\subsection{Simulating with TauRunner}

To quantify the effects of our RHN model on Earth-traversing UHE $\tau$ neutrinos, we adapted the Python-based TauRunner program \cite{Safa:2022,Safa:2020} to simulate the processes described. Using a Monte Carlo approach and numerical methods, the base version of TauRunner simulates the behaviour of such neutrinos in the Standard Model, including features such as NC and CC interactions, charged lepton energy losses, and charged lepton decay. The impact of $\tau$ regeneration, which emerges from combining these steps, is considered.

\begin{figure}
    \centering
    \includegraphics[scale=0.33]{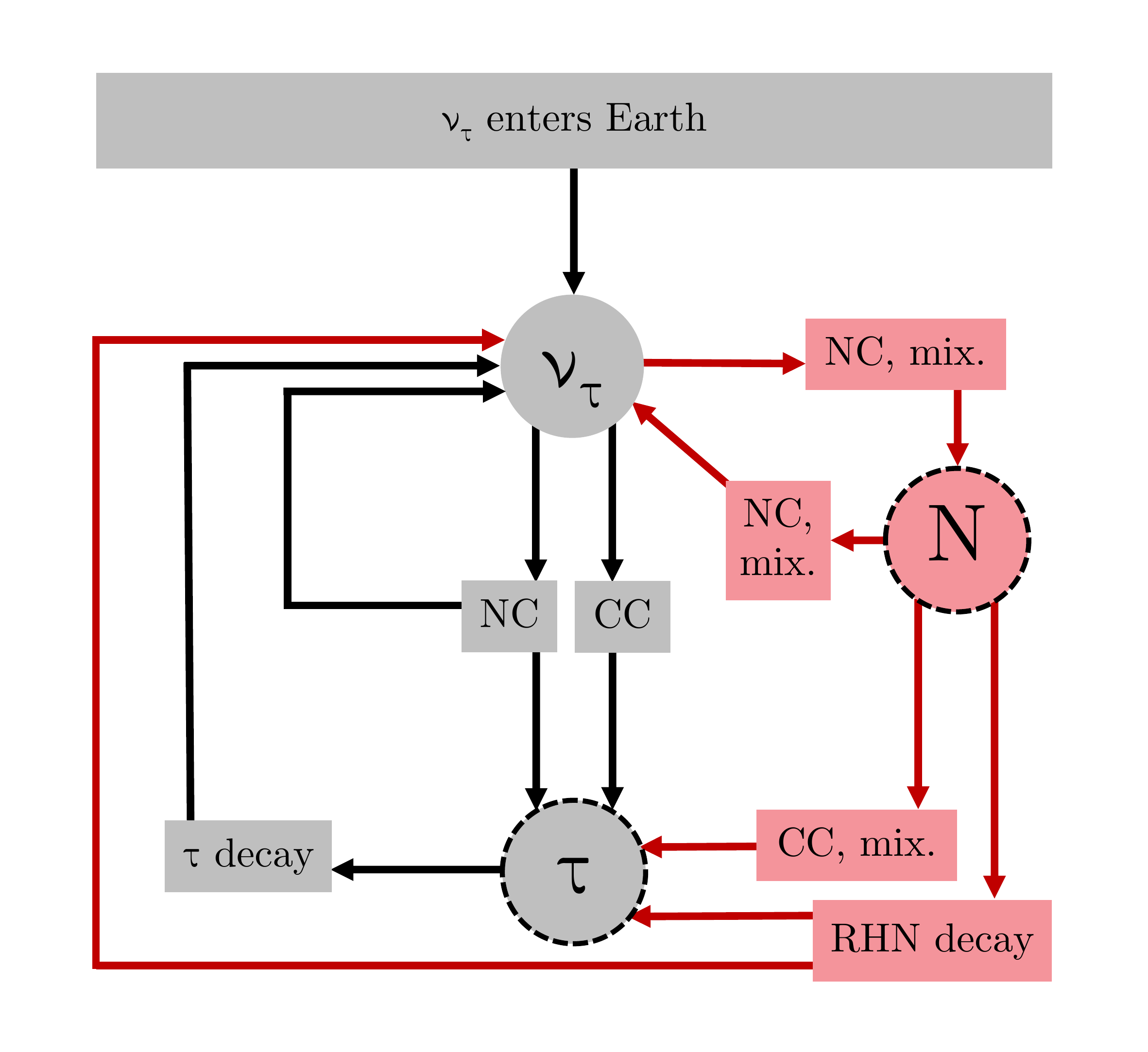}
    \caption{\label{fig:flowchart} \footnotesize Flowchart illustrating the simulation of a particle in our adapted TauRunner program. Grey items are featured in the base version of TauRunner, while red items represent the BSM modifications implemented in our adapted version. Dashed bordering is applied to those particles that, having exited the Earth, may enable detection via an EAS in the atmosphere.}  
\end{figure}

We introduce the right-handed neutrino $N$ to TauRunner's existing inventory of particle species and establish its interactions with Standard Model particles accordingly, including the new mixing NC and CC interactions and the decay of the RHN. Fig.~\ref{fig:flowchart} illustrates the possibilities open to a particle simulated by our adapted TauRunner program as it traverses a chord length through the Earth. The path choice at any given juncture is determined by Monte Carlo methods involving the random sampling of distributions derived from the relevant cross sections, decay widths, and branching ratios.

\begin{figure}
    \centering
    \includegraphics[width=\linewidth
    ]{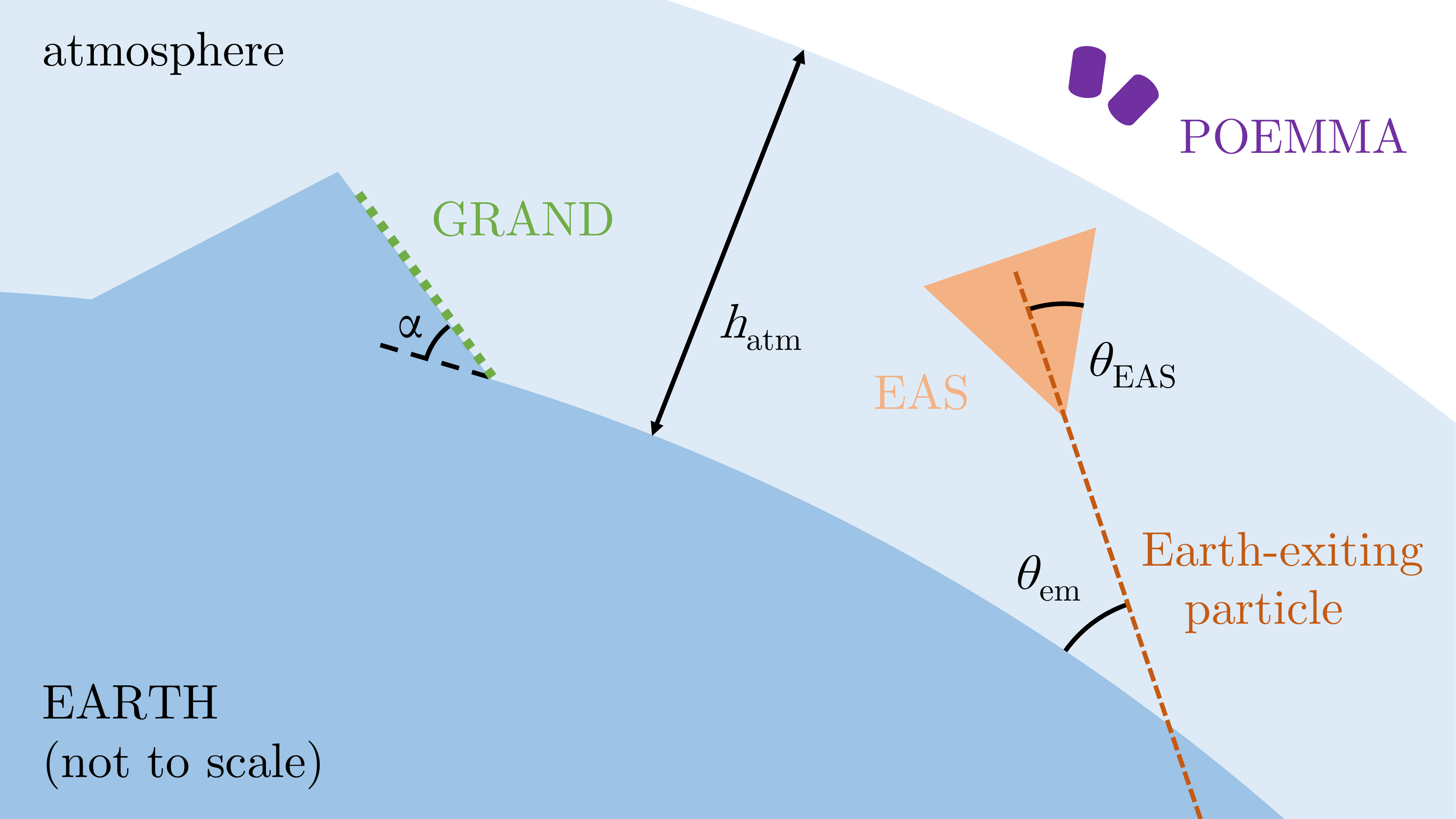}  
    \caption{\label{fig:angles} \footnotesize A schematic to illustrate the relevant geometry: the emergence angle $\theta_{\text{em}}$, the EAS angle $\theta_{\text{EAS}}$, and the atmospheric altitude $h_{\text{atm}}$. We additionally depict GRAND (with its incline of $\alpha$) and POEMMA for completeness.}
\end{figure}
For a given set of model parameters (the RHN mass $m_N$ and the mixing angle $\theta_{\text{mix}}$) and simulation specifications (the number $n$ and energy $E_{\nu}$ of the initial sample of UHE $\tau$ neutrinos, and the emergence angle $\theta_{\text{em}}$ determining the chord length they traverse through the Earth), our adapted TauRunner program provides as an output a list of all Earth-exiting particles and their energies. 

For both species of interest (the $\tau$ and the RHN), we calculate the number of exiting particles as a fraction of the initial $\nu_{\tau}$ sample, henceforth known as $P_{\text{exit}}$. It should be noted that while $P_{\text{exit}}$ can be intuitively regarded as the probability of an event resulting in an exiting particle of the given species, it does not correspond directly to a probability; one could, in principle, find that $P_{\text{exit}} > 1$ due to the production of multiple daughter particles by RHN decays, though with the parameter ranges investigated we generally expect $P_{\text{exit}} \ll 1$. 

The adapted TauRunner was run for a range of geometries between $\theta_{\text{em}}=0.1^\circ$ and $\theta_{\text{em}}=90^\circ$, where we define the emergence angle $\theta_{\text{em}}$ between an Earth-exiting particle's path and the surface of the Earth at its point of emergence (see Fig.~\ref{fig:angles} for illustrative detail). An angle of $\theta_{\text{em}}=0^\circ$ thus denotes a path tangential to the Earth's surface, with low values of $\theta_{\text{em}}$ representing Earth-skimming particles and higher values representing longer chords through the planet's interior (up to $\theta_{\text{em}}=90^\circ$, which corresponds to a path traversing the full diameter of the Earth).

\begin{figure*}
    \includegraphics[width = 0.333\linewidth]{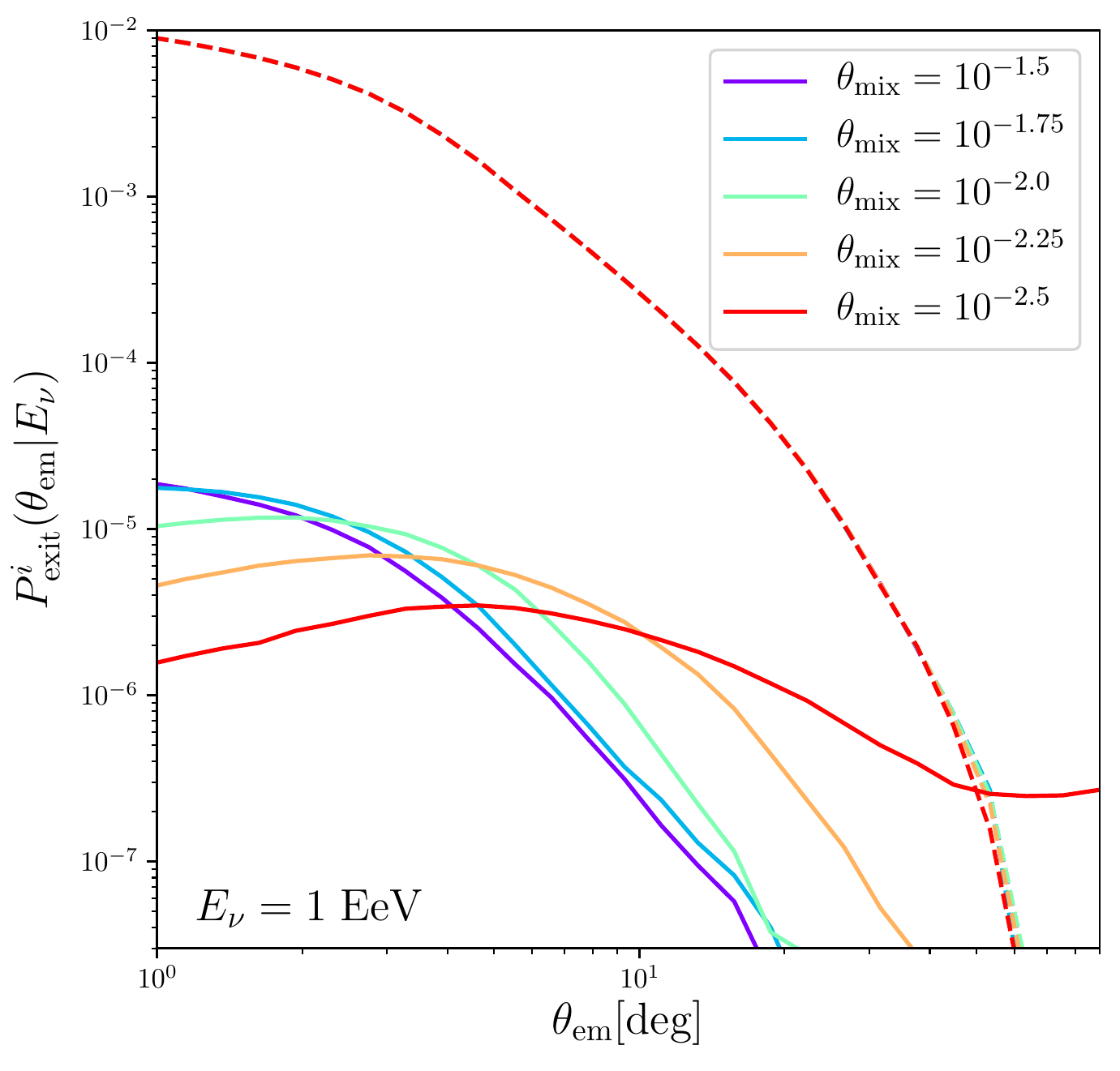}\includegraphics[width = 0.333\linewidth]{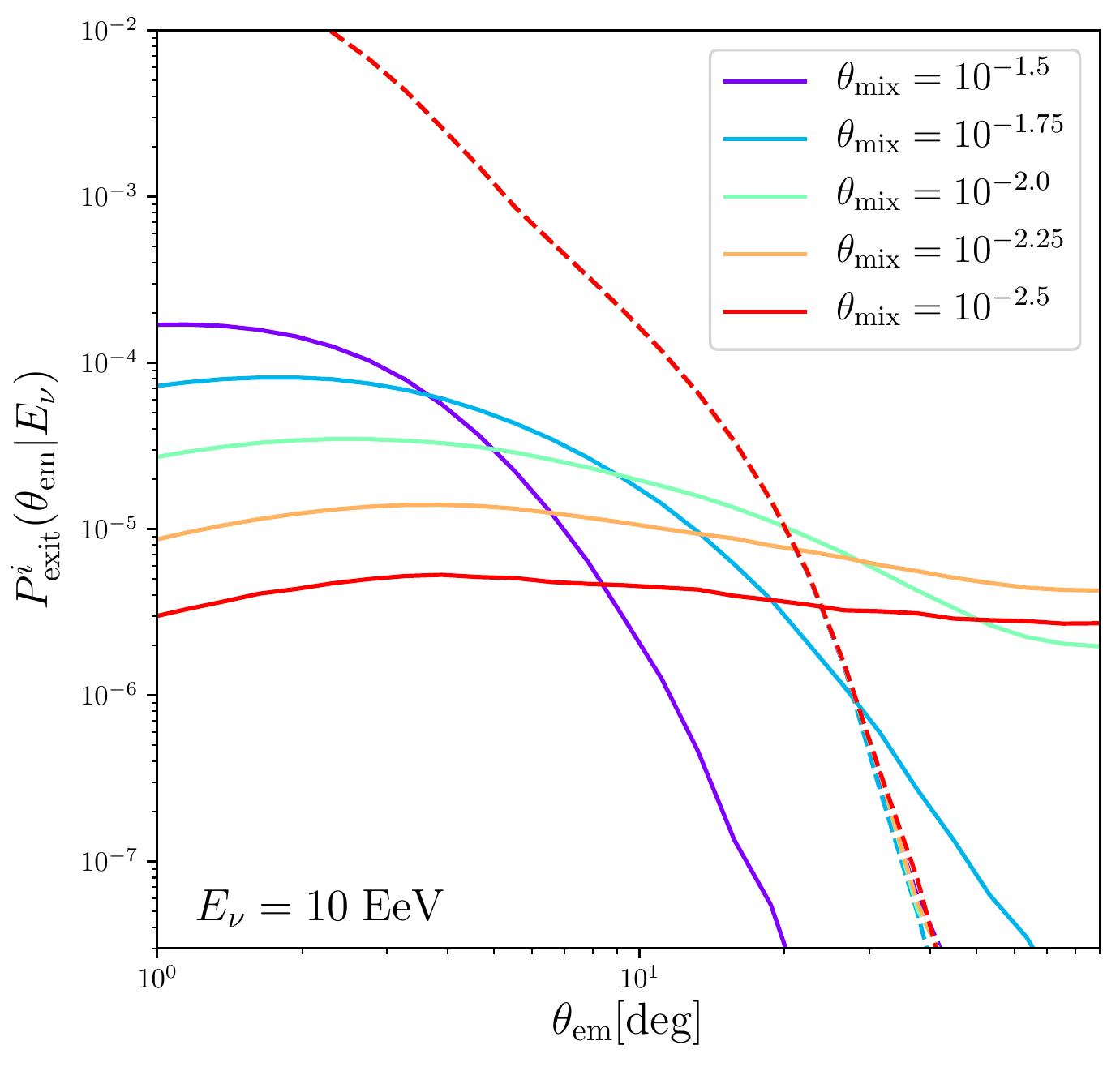}\includegraphics[width = 0.333\linewidth]{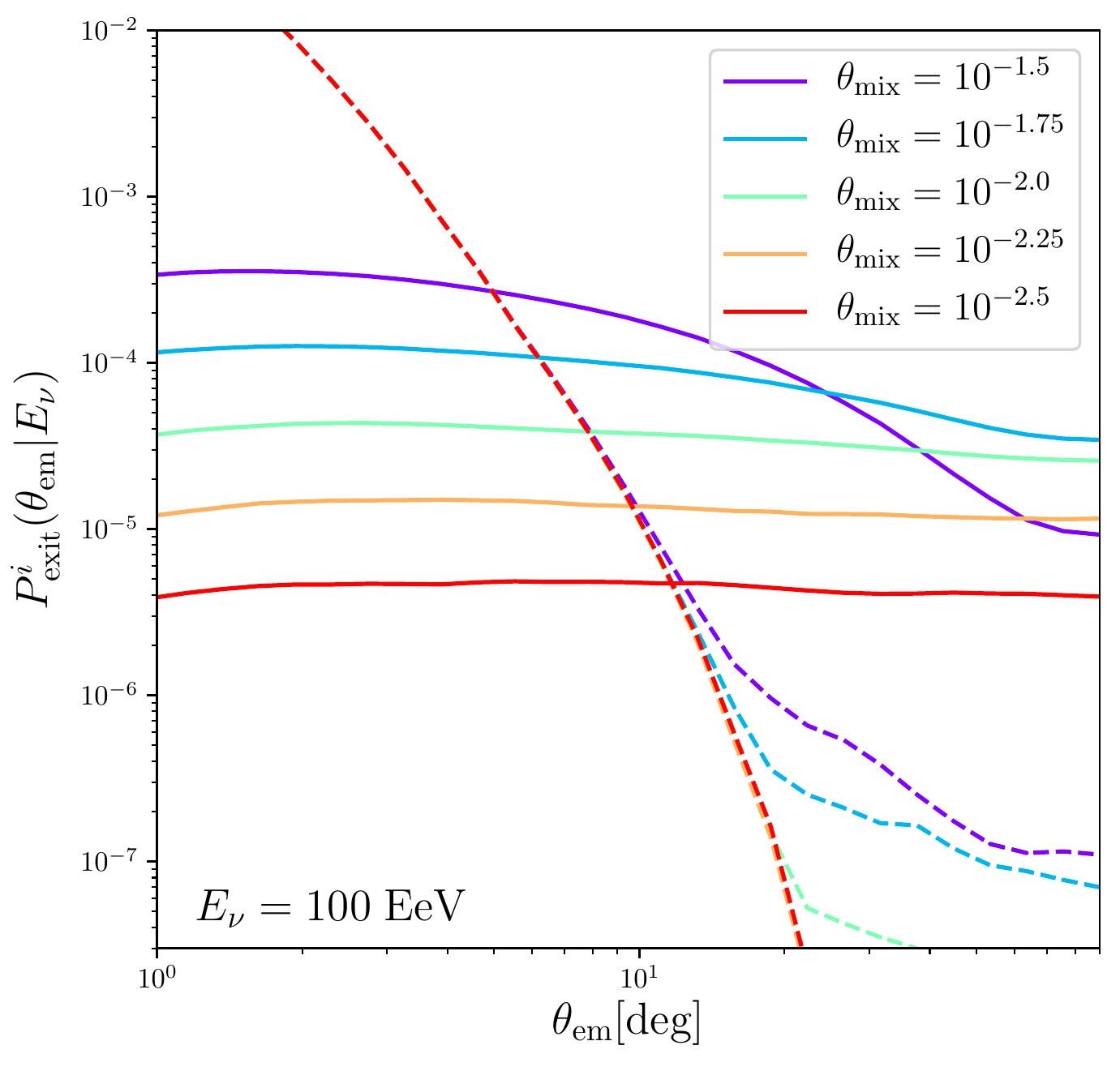}
    \caption{\label{fig:Pexit} \footnotesize Variation of $P_{\text{exit}}$ with the emergence angle $\theta_{\rm em}$, for a chosen RHN mass of $m_N = 3$ GeV and various choices of the mixing angle $\theta_{\text{mix}}$, simulated for different initial neutrino energies $E_{\nu}$.}
\end{figure*}

Fig.~\ref{fig:Pexit} shows the variation of $P_{\text{exit}}$ with $\theta_{\text{em}}$ for different choices of the parameters and initial $\nu_{\tau}$ energy $E_{\nu}$. 

At low emergence angles, corresponding to Earth-skimming events, the mixture of detectable particles is vastly dominated by the $\tau$ leptons expected in the Standard Model case. The comparatively minuscule quantity of RHNs depends on $\theta_{\text{mix}}$; a higher value of the mixing angle results in more mixing NC interactions capable of producing RHNs. For the short paths through the Earth's interior constituted by low emergence angles, this RHN production is the dominant process influencing $P^{N}_{\text{exit}}$.

At higher emergence angles, corresponding to longer chords, the Earth becomes effectively opaque to UHE neutrinos in the Standard Model scenario. As a particle traverses such a large distance through the planet's interior, it loses energy via numerous interactions until it can no longer be considered a UHE neutrino. Our adapted TauRunner program implements this via a chosen minimum-energy threshold below which a simulated particle is neglected. For the $P_{\text{exit}}$ plotted in Fig.~\ref{fig:Pexit}, this cut-off is set 2.5 orders of magnitude below the initial energy $E_{\nu}$ of the $\nu_{\tau}$ sample (e.g. for $E_{\nu}$ = 10 EeV, $P_{\text{exit}}$ is defined to include only those particles that exit with energy $E \gtrsim 32$ PeV).

While the SM (or low-$\theta_{\text{mix}}$) $\tau$ flux drops away at these large distances through the Earth, higher choices of $\theta_{\text{mix}}$ allow for a small flux to be retained, especially at higher $E_{\nu}$. The influence of the BSM physics, namely regeneration via the production, propagation, and decay of RHNs, permits the survival of more particles that may exit as $\tau$ leptons. As demonstrated in the rightmost plot of Fig.~\ref{fig:Pexit}, greater values of $\theta_{\text{mix}}$ improve this $\tau$ lepton `tail'.

The right-handed neutrino flux often dominates this high $\tau$ flux at high energies. While for Earth-skimming events (small $\theta_{\text{em}}$), the RHN flux was improved by increasing the mixing angle $\theta_{\text{mix}}$, at longer chord lengths (large $\theta_{\text{em}}$) the higher mixing angles begin to deplete it. Over these length scales, the possible decay of the RHN while still traversing the Earth becomes significant, rivalling the production of the RHN (via mixing NC interactions) in influence on the scaling of $P^{N}_{\text{exit}}$ with $\theta_{mix}$. A higher mixing angle allows for increased production of RHNs, but additionally reduces the average length $\lambda_{N}$ over which the RHN may propagate before decaying, as depicted in Fig.~\ref{fig:decay_length}. In many cases, the latter effect dominates. As a result, we may see in the leftmost plot of Fig.~\ref{fig:Pexit}, for example, that the ordering (by the size of $P^{N}_{\text{exit}}$) of the mixing angles used becomes entirely inverted as we probe higher emergence angles.

The significance of the RHN flux and BSM $\tau$ flux relative to the SM $\tau$ background at higher emergence angles renders this region of $\theta_{\text{em}}$ an area of particular interest. In later parts of this work, we especially focus on these deeper particle track geometries.

\section{From Propagation to Detection}
\label{sec:detector}
In the previous sections, we have described how the presence of new physics --- particularly of a right-handed neutrino mixing with the $\tau$-neutrino --- affects the propagation of UHECRs through the Earth. We now explore the capacity of large-volume detectors to probe the existence of new physics in the future using UHECRs. 

\subsection{Effective Area}
\label{sec:Aeff}

Let us first set up the general framework we use to compute the effective area of a detector. First of all, we will restrict our study to detectors that hunt for the production of extensive air showers (EAS) in the atmosphere after a boosted particle decays into hadrons. This regroups balloon experiments such as ANITA or PUEO, space-based observatories such as POEMMA, and Earth-based detectors such as GRAND, Trinity, or Taroge-M. For this class of detectors, once a charged or neutral lepton exits the Earth, it may lead to a detectable signal under two conditions: $(i)$ it must decay within the atmosphere for the EAS to develop fully, and $(ii)$ this EAS must feature a strong enough electric field peaking in the vicinity of the detector (for radio antennas) or a sufficiently high photon yield (for photon detectors) to trigger observation. As a result, the effective areas we aim to calculate can be expressed,  following Ref.~\cite{Heurtier:2019git}, as 
\begin{widetext}
\be \label{eq:integral}
\frac{d^3 \mathcal A^i_{\rm eff}}{d\Omega_\nu dE_\nu} = \int R_E^2 d\Omega_E \int dE_{\rm exit} \frac{dP_{\rm exit}^i}{dE_{\rm exit}}(\theta_{\rm em}|E_\nu)\int d\ell_{\rm dec} \frac{dP_{\rm dec}^i}{d\ell}(\ell_{\rm dec} | E_{\rm exit}) 
P_{\rm det}^i(E_{\rm exit},\vec{r}_{\rm dec})\,,\qquad i=\tau\,,\ N\,.\ee
\end{widetext}
In this expression, $dP_{\rm exit}^{i=\tau,N}/dE_{\rm exit}$ denotes the probability that an incoming neutrino with energy $E_\nu$, whose propagation chord exits the Earth with emergence angle $\theta_{\rm em}$, exits the Earth in the form of a $\tau$ or $N$ particle with energy $E_{\rm exit}$. Note that for a given incoming flux in the direction $(\theta_\nu, \phi_\nu)$, the emergence angle $\theta_{\rm em}$ is an implicit function of the angles $(\theta_E,\phi_E)$ with which a cosmic ray hits the Earth's surface.

After exiting the Earth, the decay probability of a particle $i=\tau,N$ with decay length $\lambda_i$ can be written as a function of its energy $E_{\rm exit}$ and the distance $\ell_{\rm dec}$ travelled since exiting the Earth,
\be \label{eq:Pdec}
\frac{dP_{\rm dec}^i}{d\ell}(\ell_{\rm dec} | E_{\rm exit}) = \frac{1}{\lambda_i(E_{\rm exit})}\exp\left(-\frac{\ell_{\rm dec}}{\lambda_i(E_{\rm exit})}\right)\,.
\ee
Finally, after the particle decays at the location\footnote{We define $\vec{r}_{\rm dec}$ as the vector with origin at the particle exit and end at its decay location.} $\vec{r}_{\rm dec}$, the probability that the event triggers the detector is defined as $P_{\rm det}^i(E_{\rm exit},\vec{r}_{\rm dec})$.

Using this generic formula, we will now compute the effective area of two qualitatively different detectors: GRAND and POEMMA. The former is a ground-based detector that covers a vast surface area \mbox{($S \sim \mathcal O(10^5)$ km$^2$)} but is located at relatively low altitude \mbox{($h\sim\mathcal O(1)$ km)}. In contrast, the latter has a much smaller spatial extension \mbox{($S \sim \mathcal O(1)$ m$^2$)} but is at a very high altitude \mbox{($h\sim\mathcal O(100)$ km)}, benefiting from a very large field of view. Although both features may play an essential role in the search for new physics, we will see that only one of the two configurations is beneficial when searching for a GeV-scale right-handed neutrino.

\subsection{GRAND}
The Giant Radio Array for Neutrino Detection (GRAND) \cite{GRAND:2018iaj} is a planned observatory for detecting UHE cosmic rays and neutrinos, constituting an array of radio antennae spread over a mountainous slope. These antennae are expected to receive the radio emissions of extensive air showers (EAS) that may result from $\tau$ and RHN decays in our scenario.

In order to model the physical arrangement of GRAND, we consider a plane inclined at an angle $\alpha$ to the ground at the detector site. We establish a semicircular region of radius $R_{\text{det}}$ within this plane, positioned such that the straight edge constituted by the semicircle’s diameter is in contact with the ground along the base of the slope. This configuration thus represents a region of detectors over an inclined area of $\pi R_{\text{det}}^2 /2$. We choose the radius of the semicircular region to have radius $R_{\text{det}}  = 80$ km in approximate accordance with GRAND’s expected detector area of 10,000 km$^2$ per site, and choose an inclination of $\alpha = 3^\circ$ to describe a realistic slope available for GRAND’s use~\cite{GRAND:2018iaj}.

For a given initial energy $E_{\nu}$, our simulation of GRAND uses the results of our adapted TauRunner program to determine $dP_{\rm exit}^i/dE_{\rm exit}$ for each particle species $i=\tau,N$, interpolated over the full range of the elevation angle $\theta_{\text{em}}$. We then follow the prescription described in Sec.~\ref{sec:Aeff}; for each exit energy bin, corresponding to an average decay length of the exiting particle, the probability of decay is calculated at successive points along the particle’s path through the atmosphere after exiting the Earth, and, at each point, it is determined whether or not such a decay would trigger a detection in any part of the detector region via the radiative cone produced by its EAS. Keeping only those points where a decay would trigger a detection, the decay probabilities are summed in order to integrate along the particle’s path (corresponding to the integration over $\ell$ in Eq.~\eqref{eq:integral}), and the results for the respective energy bins are combined (corresponding to the integration over $E_{\text{exit}}$) to calculate an overall probability of detection for any given set of angles describing the particle’s trajectory.

The simulation then iterates over varying incoming particle orientations (corresponding to a grid over the celestial sphere) and over a grid of impact locations on the Earth’s surface, thus integrating over geometric configurations to calculate a total effective area for the detector for a given initial neutrino energy $E_{\nu}$. The resulting effective area is multiplied by a factor of 20 to account for the multiple detectors intended for construction by the GRAND collaboration. 

\begin{figure}
    \centering
    \hspace{-0.07\linewidth}\includegraphics[width=\linewidth
    ]{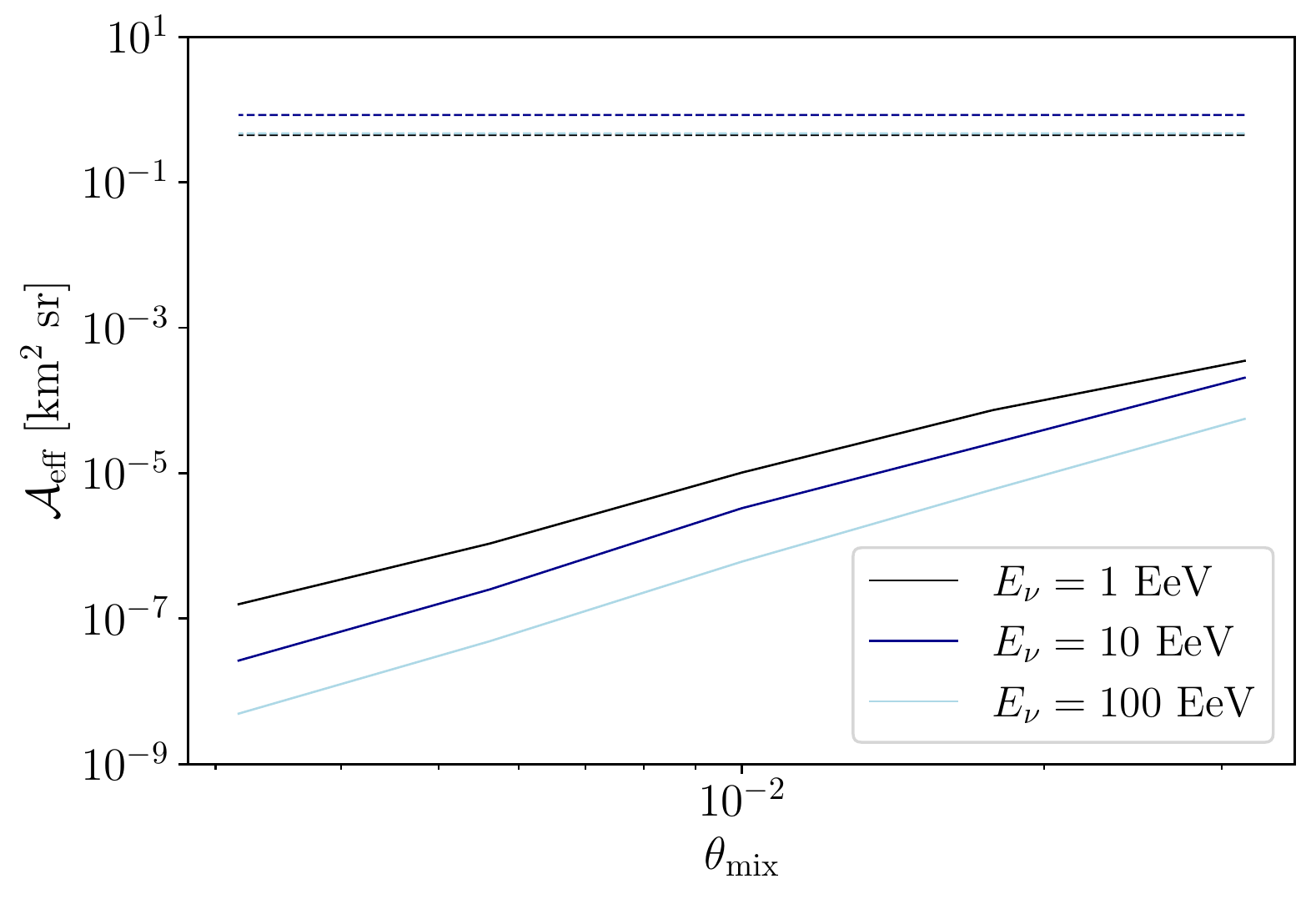}  
    \caption{\label{fig:GRAND01} \footnotesize The variation of the effective area $\mathcal A^{i}_{\text{eff}}$ calculated from our GRAND simulation with the mixing angle $\theta_{\text{mix}}$, for $\tau$ leptons ($i=\tau$, dashed) and RHNs ($i=N$, solid), for different choices of the initial neutrino energy $E_{\nu}$. We note for clarity that the dashed black line and dashed light blue line for $E_{\nu} = 1$ EeV and $E_{\nu} = 100$ EeV $\tau$ leptons respectively are overlapping.}
\end{figure}

In Fig.~\ref{fig:GRAND01} we show the results we obtain for the GRAND effective area, simulated for varying mixing angle $\theta_{\text{mix}}$, integrated over all possible incoming directions $\Omega_\nu$, and for different choices of the initial neutrino energy $E_{\nu}$. As one can see from the figure, the effective area for detecting $\tau$ leptons, $\mathcal A^{\tau}_{\text{eff}}$, is mostly insensitive to the value of $\theta_{\text{mix}}$, as any BSM effects are dominated by the SM $\tau$ flux. The predicted effective area for detecting RHNs, $\mathcal A^{N}_{\text{eff}}$, increases with $\theta_{\text{mix}}$, but is many orders of magnitude smaller than that for $\tau$ leptons, and hence cannot produce a significant signal relative to the SM background. 

\begin{figure}
    \centering
    \includegraphics[width=\linewidth]{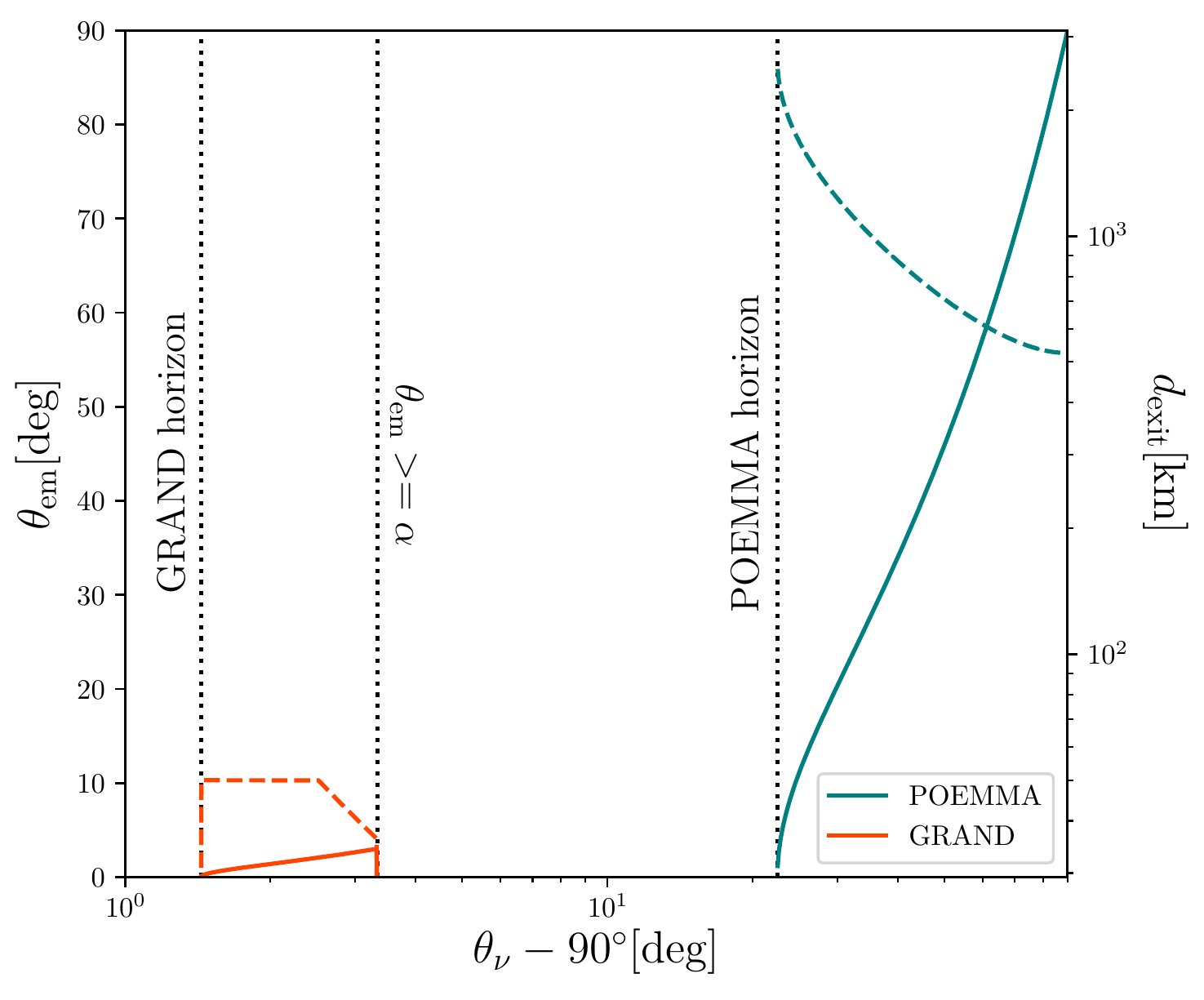}
    \caption{\label{fig:geo} Evolution of the emergence angle (plain lines, left $y$-axis) and the average distance between the detector location and the CR exit point (dashed lines, righ $y$-axis), as a function of the incoming UHECR zenith angle with respect to the detector vertical.}   
\end{figure}

This conclusion results from a combination of different constraints at play in the case of GRAND. First, the range of emergence angles accessible to GRAND is relatively narrow, limited to $\theta_{\rm em}\lesssim \mathcal O(1^{\circ})$. In Fig.~\ref{fig:geo} (left $y$-axis) we represented the evolution of the emergence angle of an UHECR arriving on Earth with a zenith angle $\theta_\nu$ (placing the zenith at the location of a GRAND detector). This emergence angle is mainly limited by the low inclination of the detector. Indeed, for a mountain slope of elevation $\alpha$, it is clear that GRAND cannot observe UHECRs emerging with angle $\theta_{\rm em}\lesssim \alpha$ (the only margin of error lying in the opening angle of a Cherenkov cone). At such low emergence angles, SM $\tau$ particles easily propagate, whereas RHNs are difficult to produce. Moreover, the range over which the exiting particle has to decay to create an EAS that is detected by GRAND is limited. As can be seen from the left $y$-axis of Fig.~\ref{fig:geo}, the typical distance between the exit point and the detector grows with decreasing emergence angle, but is quickly limited by the fact that the detector faces other mountains, about 50 km away, that act as a cut-off on the propagating chord through the atmosphere at low emergence angles.

For an isotropic GZK spectrum, we found, using the current limits from Auger~\cite{PierreAuger:2019ens}, that even a detector with $\alpha=30^\circ$ (which is far beyond realistic values implementable on Earth), GRAND would only detect an $\mathcal O(1)$ number of BSM events after 5 years of exposure when integrating over all emergence angles and exit energies. As can be seen from Fig.~\ref{fig:GRAND01}, this number of events is larger at lower energies, since the effective area increases with energy. This suggests that if cuts were applied to the exit energies and emergence angles, and bright enough transient sources were considered, GRAND may be able to obtain a measurable signal arising from the existence of RHNs. However, we could not find a realistic situation where this happens. Indeed, there are several factors at play, on top of the limitations already mentioned: 

First, in order to efficiently observe transient events, the field of view of GRAND would need to be quite precisely aligned with the incoming direction of the burst. However, the largest version of GRAND would be made of 20 separate detectors that are likely to point in different directions, significantly reducing the effective area for a given transient event.
In addition, as can be seen from Fig.~\ref{fig:Pexit}, the lowest energy channel sees BSM events only dominate at very large emergence angles that are much larger than those accessible to GRAND.

We hence find that GRAND is inappropriate for probing our particular BSM model. Due to its geometry, GRAND is expected to primarily observe events with low emergence angles, corresponding to particles that skim the Earth with short chord lengths. As illustrated by the variation of $P_{\text{exit}}$ with $\theta_{\text{em}}$ in Fig.~\ref{fig:Pexit}, neither the production of RHNs nor the variation in detectable $\tau$ flux are expected to produce a significant signal relative to the SM $\tau$ background at these low emergence angles, and so, for this BSM scenario, a detector capable of observing events at higher emergence angles (corresponding to longer chords through the Earth) is of greater interest.

We wish to emphasize that, although GRAND is disfavored compared to other space-based telescopes in this particular situation, it could be competitive in searching for BSM scenarios that involve particles with shorter lifetimes than our RHN candidate. Models where an extended spectrum of heavy particles with decay length $\lesssim\mathcal O(1-10)$ km (such as string theory or models involving extra dimensions, see e.g.~\cite{Dienes:2011ja, Dienes:2021woi, Pilaftsis:1999jk, Dienes:1998sb}) could potentially lead to situations where a large number of BSM particles exit the Earth and decay within a short distance. While such a possibility may be promising for UHECR searches using large-volume detectors such as GRAND, we leave such a study for future work. 

\subsection{POEMMA}
The Probe Of Extreme Multi-Messenger Astrophysics (POEMMA) comprises two satellites orbiting at an altitude of about $525~\mathrm{km}$ \cite{Olinto:2017xbi}. It is designed to operate using two possible configurations: one using the stereo air fluorescence technique, where the satellites are positioned in a quasi-nadir viewing configuration (POEMMA-stereo mode), and the other targeting upward-going tau neutrinos via Cherenkov signals where the two satellites are pointed closer to the Earth limb (POEMMA-limb mode). The POEMMA instruments are programmed to point towards the direction of a transient source rapidly and can track it over time.

Given the very high altitude of the detector, as compared to the atmospheric location where the EAS is emitted, POEMMA will benefit from an extraordinary field of view. As can be seen from Fig.~\ref{fig:geo}, it can probe both very small and very large emergence angles depending on the mode chosen, and the distance $d_{\rm exit}$ between the exit point of UHECRs and the detector is significantly larger than for GRAND, providing plenty of room for long-lived particles to decay before reaching the detector. Another advantage of this high altitude is that the extension of the Cherenkov cone at the detector level may spread over a hundred kilometres, thus greatly enhancing the detector's effective area despite its small spatial extension compared to GRAND. Nonetheless, the photon yield must be sufficiently large at the detector level for an EAS to trigger detection. Accurately computing the detection probability $P_{\rm det}^i$ would therefore require simulating with precision the development of the shower within the atmosphere, estimating the flux of photons arising from the EAS propagation, and running a proper detector simulation. Performing such a thorough analysis is beyond the scope of this paper, as it would require using the detector simulation programs developed by the POEMMA collaboration itself to derive robust predictions. Instead, we would like in this paper to demonstrate that such an analysis is worth the effort by using simplified but realistic approximations to obtain POEMMA's effective area. In particular, we will consider for each exiting particle that its decay triggers detection, except for
\begin{enumerate}
    \item Events for which the detector is not contained within the Cherenkov cone over which the EAS spreads;
    \item Exiting particles that decay outside the lower atmosphere, for which the EAS would not fully develop~\cite{Romero-Wolf:2018zxt,Reno:2019jtr,Cummings:2020ycz};
    \item Events originating from the decay of particles with exit energies $E_{\rm exit}<0.1~\mathrm{EeV}$ (for reasons that will become clear later). 
\end{enumerate}
Before continuing, we make a few comments to clarify how we implemented some of these conditions numerically. Indeed, both the capacity of a decaying particle to create an EAS and the opening angle of the resulting Cherenkov cone vary continuously with the altitude at which the decay occurs, and also with the energy and direction of the particle decaying. Moreover,  once the shower is created, its density of photons per square metre at the level of POEMMA can be attenuated if it has to cross a thick layer of atmosphere on the way. An extensive description of the different elements that affect the detection of an EAS, and the capacity of POEMMA to detect such EASs, can be found in Refs.~\cite{Romero-Wolf:2018zxt,Reno:2019jtr,Cummings:2020ycz}. From Ref.~\cite{Cummings:2020ycz}, which presents numerical results in the case of an emergence angle $\theta_{\rm em} = 10^\circ$, one can observe that decays at high altitudes ($\gtrsim 20$ km) may lead to EASs with a reduced opening angle accessible to detection by POEMMA ($\lesssim 0.5^\circ$). In contrast, decays at lower altitudes ($\lesssim 15$ km) can lead to an opening angle as large as $\mathcal O(3^\circ-5^\circ)$. To avoid reproducing the results of Ref.~\cite{Cummings:2020ycz} numerically, we assumed a fixed value of the opening angle detectable by POEMMA, choosing benchmark values that spread within the range $1.5^\circ\leqslant \theta_{\rm EAS}\leqslant 3.5^\circ$. The altitude at which the EAS is produced also affects the efficiency of the detection~\cite{Cummings:2020ycz}. Throughout this work, we fix the maximum altitude for the production of the EAS, and survey over several benchmark values within the range $15\mathrm{km}\leqslant h_{\rm EAS}\leqslant 25\mathrm{km}$ to estimate the influence of such a choice on the final results\footnote{From Refs.~\cite{Krizmanic:2020shl, Kakimoto:1995pr}, one can also observe that the air fluorescence yield evolution with the shower's altitude is about half its peak value at about 25km, which comforts us regarding the choice of benchmark points used throughout this work.}.

\section{Searching Strategy}\label{sec:strategy}

Now that we have described how the existence of an RHN may affect the propagation of UHECRs through the Earth and their contribution to the production of EASs in the atmosphere, we shall demonstrate how the study of the angular distribution of such events may be used to test such a BSM scenario.

From the probability distributions exhibited in Fig.~\ref{fig:Pexit}, it is clear that an RHN can affect the capacity of UHECRs to exit the Earth's surface, especially at large emergence angles. To test the existence of this RHN using large volume detectors such as POEMMA, we shall estimate to what extent different UHE neutrino sources may lead to a measurable deviation from the SM when detecting events at very large emergence angles. In order to do so, one requires information on the incoming flux being tested and robust theoretical predictions for the angular distribution of events expected both in the SM and in the hypothesis of a BSM scenario. 

As we have seen in the previous sections, the UHECR exit probability is insensitive to the presence of the RHN at low emergence angles $\lesssim \mathcal O(10^\circ)$ and is largely dominated by exiting UHE $\tau$'s, whose detection probability is also insensitive to our BSM hypothesis. For a given source, detecting events at low emergence angles can thus help accurately estimate the flux and direction of the incoming neutrinos considered. Thanks to the multiplicity of detectors that look for UHECRs at low emergence angles, it is therefore possible that several detectors could observe the same flux at different emergence angles. In the case of POEMMA, it would even be conceivable that the two satellites point simultaneously at the same source from different angles. In what follows, we will hence assume that, based on measurements at low emergence angles, the flux of incoming UHE neutrinos is known, regardless of the nature of the source considered. We will then study the possibility that POEMMA can observe such a source at a large emergence angle $\gg 10^\circ$, and estimate the region of parameter space that POEMMA could help probe in the future.

\begin{figure}
    \centering
    \includegraphics[width=\linewidth]{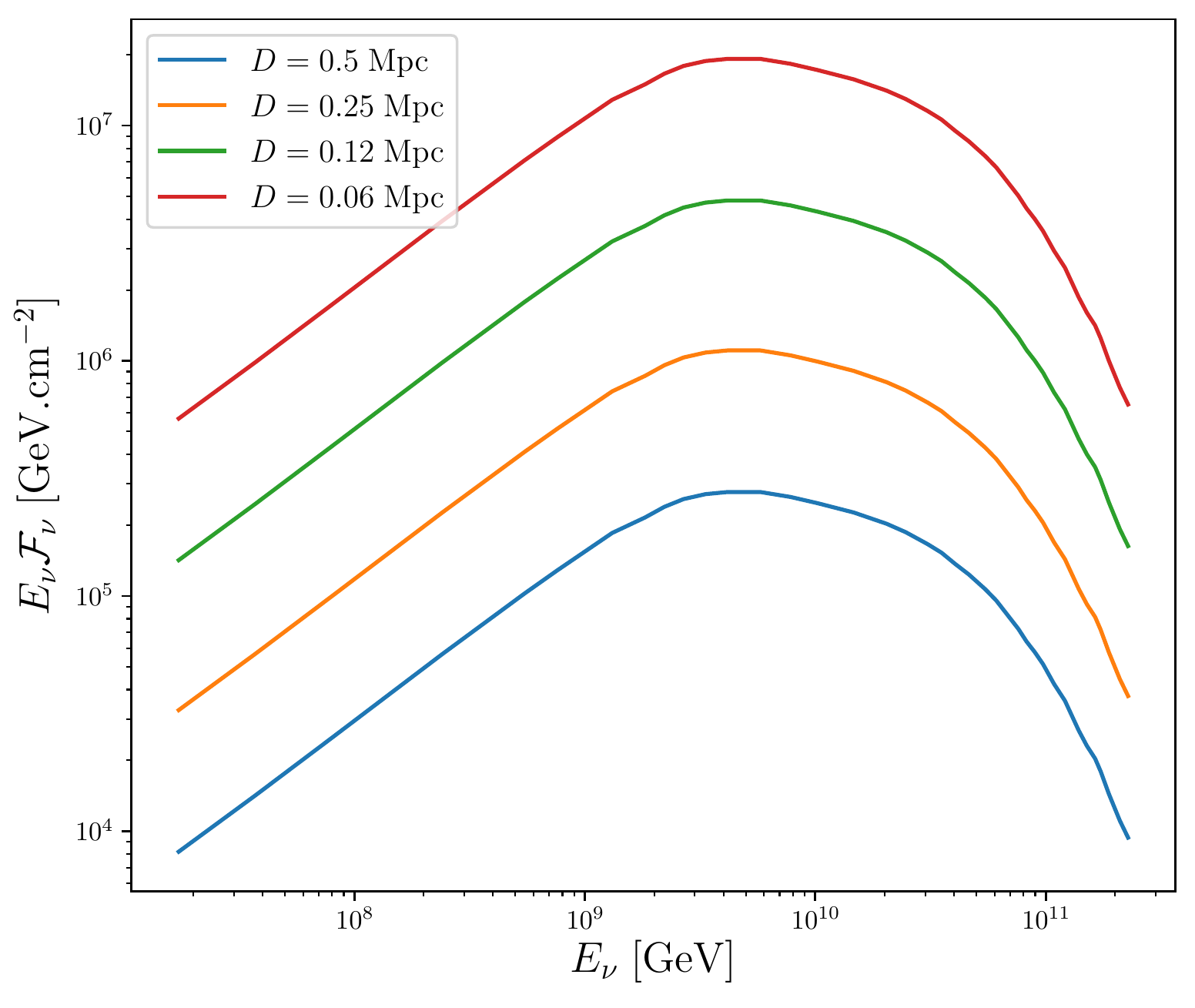}
    \caption{\label{fig:fluence}\footnotesize Neutrino fluence expected from the explosion of a GRB similar to GRB 221009A, as taken from Ref.~\cite{Rudolph:2022dky} and rescaled as a function of the distance $D$ considered.}
\end{figure}
\subsection*{\bf Diffuse vs Transient Sources}

Many different sources can produce a flux of UHE neutrinos in the Universe. Depending on the nature of the source, such a flux may be constant and isotropic (as is the case for cosmogenic (GZK) neutrinos) or may result from a transient event, localised in space, and lasting only a finite amount of time.

 The case of a diffuse UHE neutrino background is quite constrained, according to the ANITA, Auger, and IceCube collaborations\cite{PierreAuger:2019ens, ANITA:2018vwl, IceCube:2018fhm}. Given the planned sensitivity of POEMMA, even by increasing its field of view up to 360$^\circ$,  the collaboration can only hope to detect $\mathcal O(10-100)$ events in 5 years of observation~\cite{POEMMA:2020ykm}. Such events are likely to be observed preferentially at low emergence angles. Therefore, we can confidently rule out the possibility that a diffuse UHE neutrino flux helps detect the existence of new physics using large-volume detectors at large emergence angles. Instead, we will focus on what follows on UHECRs produced by transient sources.

In recent publications, POEMMA was shown to constitute an excellent instrument for performing Target-of-Opportunity neutrino observations~\cite{Venters:2019xwi}. In particular, it can detect over a hundred events when looking at short and long bursts producing UHE neutrinos at a distance of $\mathcal O(10)$ Mpc. Moreover, in the circumstance that a transient astrophysical event producing UHE neutrinos should take place at a shorter distance, this number could be easily increased by several orders of magnitude, given the scaling of the corresponding neutrino flux with the inverse distance squared.

From Fig.~\ref{fig:Pexit}, one can see that an incoming UHECR with energy $100$ EeV can produce RHN exiting the Earth at large emergence angles with probability $\mathcal O(10^{-4})$. A burst for which POEMMA could detect $10^4$ events at a low emergence angle (corresponding to a source located about an Mpc away from the Earth) could thus potentially produce $\mathcal O(1)$ event at emergence angles $\gtrsim 10^\circ$.

\begin{figure*}
    \centering
    \includegraphics[width=0.499\linewidth]{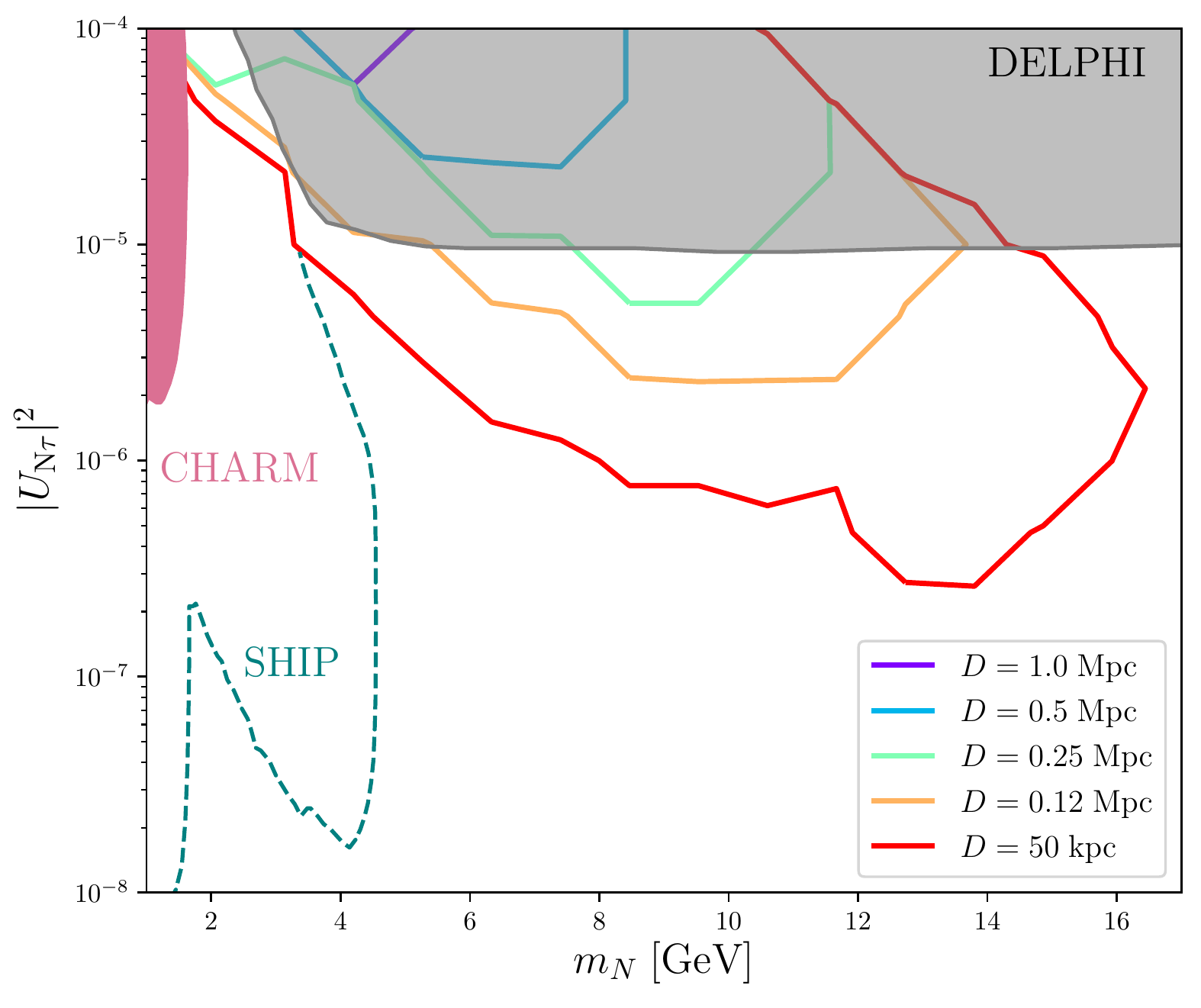}\includegraphics[width=0.499\linewidth]{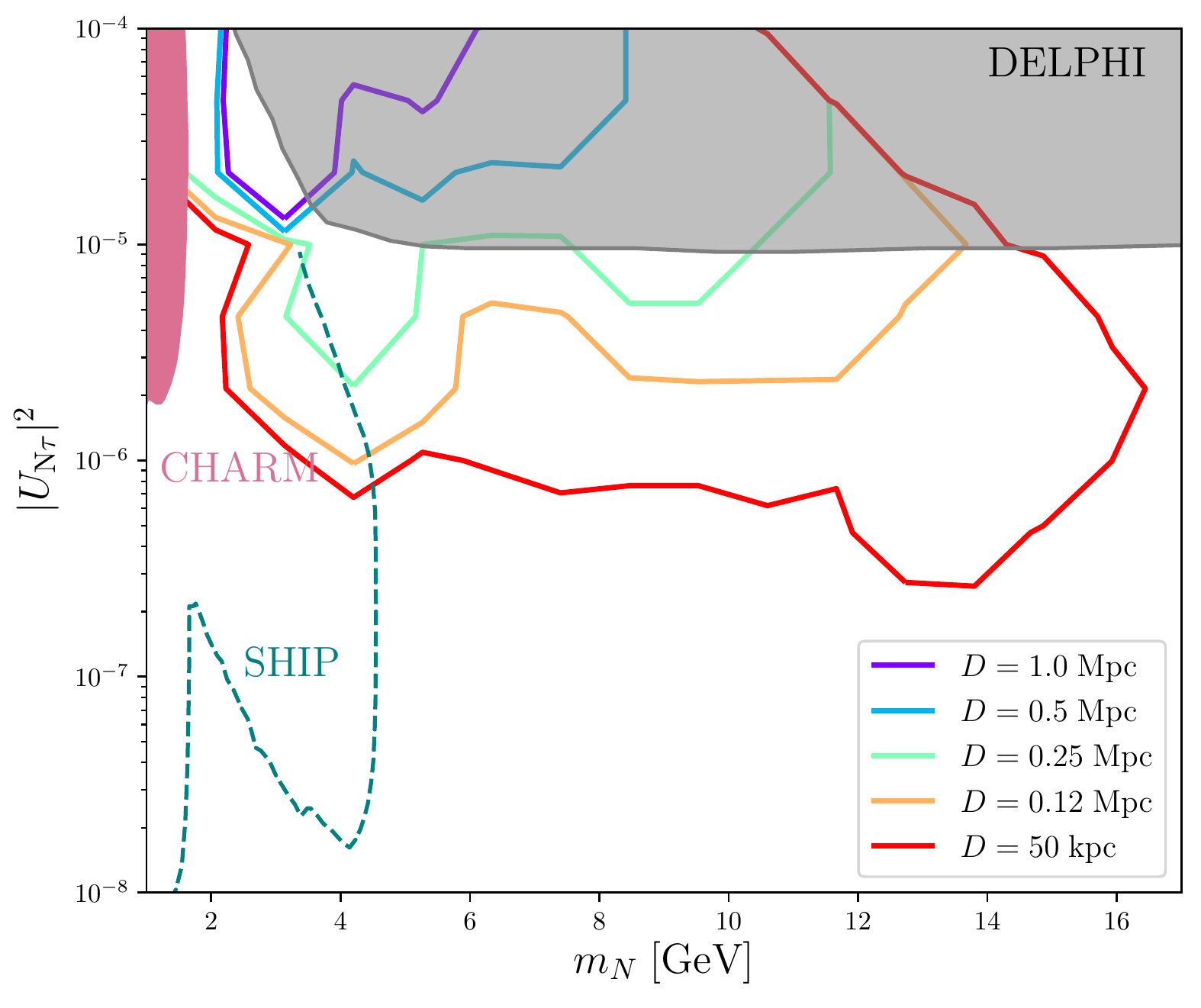}
    \caption{\label{fig:energycut} Sensitivity of the proposed RHN search using POEMMA, for an emergence angle of $60^\circ$, and considering a GRB similar to GRB 221009A, as taken from Ref.~\cite{Rudolph:2022dky} and rescaled as a function of the distance $D$ considered. The search is restricted to events with exiting energies $10\leqslant E_{\rm exit}\leqslant 100$ EeV (left panel) and $1\leqslant E_{\rm exit}\leqslant 100$ EeV (right panel).}
\end{figure*}
Gamma-ray bursts (GRBs) --- UHE explosions involving a massive star's collapse or the merger of two compact objects --- constitute prime targets of multi-messenger astronomy that can typically last over tens of tens of milliseconds to hundreds of seconds. They have been detected using gravitational waves~\cite{LIGOScientific:2017vwq} and were proposed as potential sources of UHECRs and astrophysical neutrinos (see e.g. \cite{Waxman:1995vg,Vietri:1995hs}). On October 9th 2022, a particularly bright GRB (named GRB 221009A) was observed about 637 Mpc away from the Earth~\cite{2022GCN.32648....1D}. The burst triggered the Gamma-Ray Burst Monitor (GBM)~\cite{2022GCN.32636....1V}, the Burst Alert Telescope (BAT)~\cite{2022ATel15650....1D}, the Fermi Large Area Telescope (LAT)~\cite{pillera2022grb}, and the LHAASO collaboration~\cite{2022GCN.32677....1H} at various energies. Various publications (see e.g. Refs.~\cite{Rudolph:2022dky} and \cite{AlvesBatista:2022kpg}), while trying to understand why no muon-neutrino track was detected by IceCube~\cite{2022GCN.32665....1I}, proposed models to explain the timing and energy distribution of the gamma rays observed while predicting the expected flux of UHE neutrinos up to energies of $\mathcal O(100) $ EeV.

In what follows, we will take GRB 221009A as a benchmark example and use the corresponding UHE neutrino fluence derived in Ref.~\cite{Rudolph:2022dky} to search for our RHN candidate using POEMMA. We will then extrapolate this result to sources at arbitrary redshift by considering that the UHE neutrino flux of a source at a distance $D$ from the Earth can be obtained from the flux calculated for GRB 221009A at a distance $D_{\rm GRB 221009A}$ by simply rescaling the result published in Ref.~\cite{Rudolph:2022dky} by a factor $(D_{\rm GRB 221009A}/D)^2$. We show in Fig.~\ref{fig:fluence} the energy fluence $E_\nu\mathcal F_\nu$ expected for such event as a function of the incoming neutrino energy $E_\nu$ for variance cosmic distances. In what follows, we will consider distances that scale roughly between the size of the Milky Way $~30$kpc and our distance from the Andromeda galaxy, located at about 765kpc. Although the explosion of GRBs at such distances is expected to be rare, detecting such events with UHECR detectors would constitute a compelling instrumental test for new physics.

\section{Results}\label{sec:results}
We now turn to present our results. Using the incoming $\tau$-neutrino fluence introduced above, we imagine that POEMMA could point at the event rapidly at an angle large enough to be sensitive to the effect of new physics. Simultaneously, we envision that other large-volume detectors operating at low emergence angles may detect the same event and be able to evaluate the total value of its neutrino flux. Given the value of this flux (corresponding to a given brightness and distance of the event from Earth), we used our code to calculate the energy distribution of EAS events that POEMMA could detect at various emergence angles, both in the case of the SM and in the hypothesis of an RHN that couples exclusively to the $\tau$ neutrino. For emergence angles $\gtrsim 60^\circ$, we tested with our MC simulation that the SM background is zero with a precision of $\mathcal O(10^{-10})$ for all exit energy bins considered. We hence used an exit probability for SM particles of $10^{-10}$ for these large emergence angles. Using the flux described above, we then calculated a corresponding number of events for both SM and BSM particles. We demanded a rejection of the BSM hypothesis with a 99.7\% CL.

In Fig.~\ref{fig:energycut}, we present our results in the case of an emergence angle of $60^\circ$, while restricting our search to EAS with energies $E\in [10,100]$ EeV (left panel) and  $E\in [1,100]$ EeV (right panel). As one can see from these figures, different RHN masses and mixing angles $|U_{N\tau}|\approx \theta_{\rm mix}$ can give rise to detectable signals with different exit energies. Namely, low RHN masses ($\lesssim 3$ GeV) with small mixing angles ($|U_{N\tau}|^2\sim 10^{-6}$) can be probed exclusively by selecting events with exit energy $\lesssim 10$ EeV whereas larger RHN masses can be probed down to $|U_{N\tau}|^2\sim 10^{-7}$ by searching for events with exit energy in the range $10\leqslant E_{\rm exit}\leqslant 100$ EeV. In these figures, we also indicate for comparison current limits from CHARM~\cite{CHARM:1985anb}, DELPHI~\cite{DELPHI:1996qcc}, and SHiP~\cite{SHiP:2018xqw}. Such constraints would also be complementary with future HL-LHC searches~\cite{Cheung:2020buy}, and with future dedicated UHECR searches proposed in Ref.\cite{Fischer:2023bfn} for probing the existence of HNLs in the mass range $\mathcal O(10 \mathrm{MeV}- 2 \mathrm{GeV})$.

As one can see, the sensitivity of POEMMA to the existence of an RHN is thus competitive with existing constraints, even for GRBs that would take place at distances larger than a Mpc. It may provide complementary probes to such a scenario compared to future long-lived particle search experiments.

\begin{figure*}
    \centering
    \includegraphics[width=0.499\linewidth]{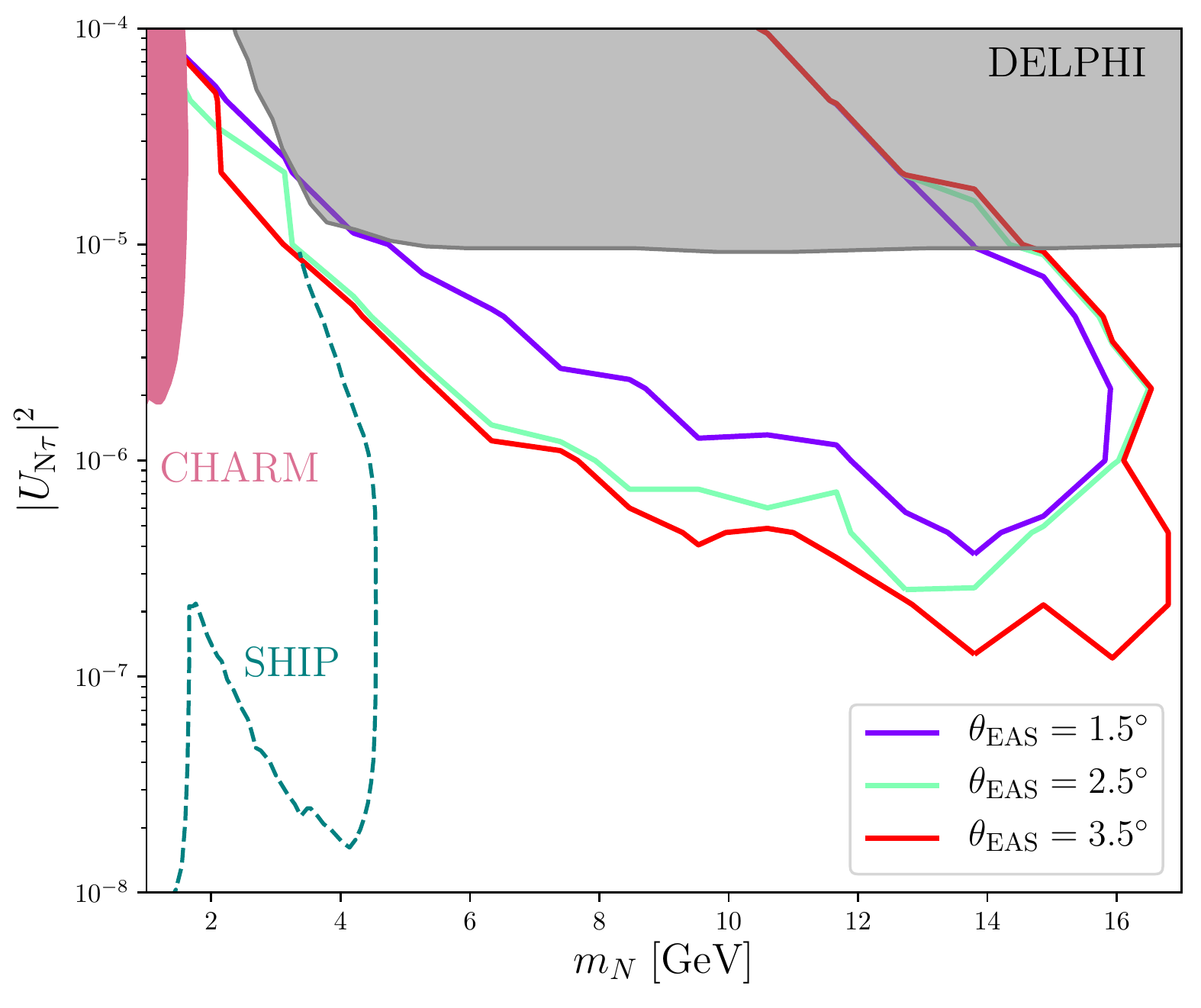}\includegraphics[width=0.499\linewidth]{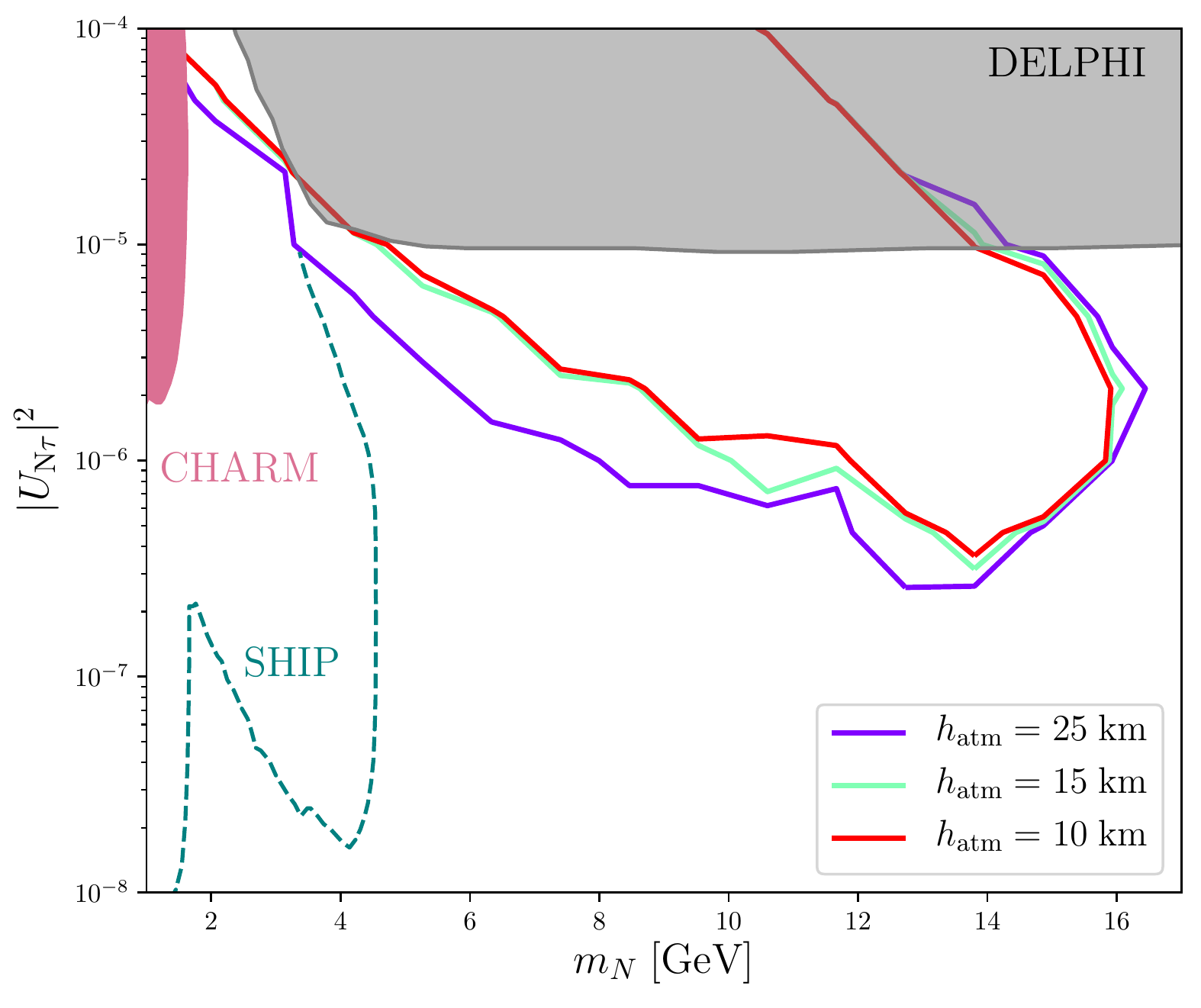}
    \caption{\label{fig:param}\footnotesize Sensitivity of the proposed RHN search using POEMMA, for an emergence angle of $60^\circ$, considering a transient source similar to GRB 221009A, as taken from Ref.~\cite{Rudolph:2022dky}, happening at $D=50$ kpc away from the Earth. The search is restricted to events with exiting energies $10\leqslant E_{\rm exit}\leqslant 100$ EeV. It uses varying values of the effective Cherenkov opening angle $\theta_{\rm EAS}$ (left panel) and atmosphere altitude $h_{\text{atm}}$ (right panel).}
    
\end{figure*}

\subsection*{Sensitivity to the Used Parameters}

As we discussed in Sec.~\ref{sec:detector}, to simulate the capacity of POEMMA to detect EASs after they are produced, we chose to fix:
\begin{itemize}
    \item The value of the larger altitude $h_{\rm atm}$ under which these showers need to form to be detectable;
    \item The effective opening angle of the Cherenkov cone $\theta_{\rm EAS}$ that can be detected by POEMMA.
\end{itemize}
Naturally, one could wonder to which extent our results are sensitive to such choices. In Fig.~\ref{fig:param}, we vary such parameters independently within the ranges motivated in Sec.~\ref{sec:detector}, for a fixed emergence angle $\theta_{\rm em}=60^\circ$ and exit energies restricted to the range $10\leqslant E_{\rm exit}\leqslant 100$ EeV. As expected, the larger the Cherenkov cone opening angle (left panel) and the larger the atmosphere altitude (right panel) considered, the more sensitive the proposed UHECR search is to our BSM hypothesis. However, it is remarkable that even by restricting the atmosphere of 10km of altitude, or the Cherenkov opening angle to 1.5$^\circ$, the search for an RHN for sources as far as an Mpc away from the Earth still allows probing a significant fraction of the parameter space.

\section{Conclusion}\label{sec:conclu}
In this work, we have studied the influence of BSM physics on the behaviour of UHE neutrinos within the Earth and their possible measurement at large-volume observatories. 

In particular, we have introduced a right-handed Majorana neutrino that mixes with the left-handed $\tau$ neutrino,
and we have presented the relevant physics affecting Earth-traversing events in this case. 
The TauRunner program, which simulates the propagation of UHE neutrinos and charged leptons through the Earth, was adapted to our model and used to compute the probability of UHECRs exiting the Earth after propagating through its interior. 

We have then considered the production of extensive air showers from the decay of $\tau$ leptons and right-handed neutrinos of GeV-scale mass. Finally, we have modelled the ground-based GRAND radio array and the space-based POEMMA observatory to estimate the detection probability of such events in the near future. 

We found that UHECRs entering and exiting the Earth with a low emergence angle (corresponding to the short chords traversed by Earth-skimming particles) produce a subdominant BSM signal in comparison to the Earth-exiting $\tau$ flux expected in the Standard Model; this severely limits the potential of GRAND, whose field of view favours the observation of small emergence angles, in this particular investigation. POEMMA, which probes larger emergence angles corresponding to longer chords through the Earth, is of greater interest in this scenario.

We have investigated the possibility of testing our BSM scenario with diffuse and transient UHE neutrino sources, such as cosmogenic GZK neutrinos or neutrinos produced during GRBs observed in nearby galaxies. We conclude that, while diffuse fluxes were too low for extracting a BSM signal over the SM background, transient sources may provide large enough fluxes and feature high enough UHE neutrino energies to allow for the identification of significant deviations from the SM case using POEMMA in the future.

Using recent simulations of the expected flux of UHE neutrinos predicted for the event GRB 221009A, we have assumed that a similar GRB would take place at various distances from the Earth and searched for regions of the parameter space where one could extract a deviation from the SM case at the 99.7\% confidence level. We have therefore scanned over RHN masses and mixing angles and identified currently unconstrained regions of the parameter space for which this can be the case.

Because the existence of BSM physics affects the propagation of UHECRs through the Earth, and thus their angular and energy distribution once they exit the Earth, we found that restricting searches of EAS that reach the detector to different ranges of energies and emergence angles can lead to probing very different regions of the parameter space. 

In particular, we have presented results for an event observed with an emergence angle of 60$^\circ$ and showed that low RHN masses ($\lesssim 3$ GeV) with small mixing angles ($|U_{N\tau}|^2\sim 10^{-6}$) could be probed exclusively by selecting events with exit energy $\lesssim 10$ EeV, whereas larger RHN masses can be probed down to $|U_{N\tau}|^2\sim 10^{-7}$ by searching for events with exit energy in the range $10\leqslant E_{\rm exit}\leqslant 100$ EeV.

We have also varied the parameters used in our detector simulations --- namely, the maximum altitude before which an EAS can be produced, and the EAS opening angle that can trigger the detector --- and have found that such a search would remain competitive regardless of the parameters chosen.

To our knowledge, this study is the first that considers the possibility of testing the existence of long-lived BSM particles using large-volume detectors. We believe this direction has a rich untapped potential that can be studied in various contexts.

In this work, we have considered that our RHN mixes exclusively with the $\tau$ active neutrino. A natural extension of this model would be to include three flavours of right-handed neutrinos that couple to all three flavours of active left-handed neutrinos and to simulate the propagation of leptons of different flavours through the Earth. Including muons in the inventory of detectable particles may be of particular interest, as the profile of UHECRs that can exit the Earth in their case has been shown to be fairly different to that for $\nu_\tau$ CRs~\cite{Cummings:2020ycz}. Also, tracking the propagation of $\nu_e$ and $\nu_\mu$ through the Earth may lead to secondary contributions to $\tau$-induced EASs that are not accounted for in this study but could be of interest for future searches~\cite{Soto:2021vdc}.

In principle, any theoretical setting in which Earth-traversing UHECRs produce relatively long-lived intermediate particles could be of interest, perhaps including theories featuring majorons, additional right-handed neutrinos, or a more extensive BSM sector. In particular, we would like to stress that a scenario in which intermediate BSM particles are abundantly produced, but have a shorter decay length as compared to our RHN candidate, may be more compelling for detection with lower altitude detectors than POEMMA, such as GRAND, Taroge-M, PUEO, or Trinity.

Another line of investigation underway is the application of such simulations to objects in the Solar System other than the Earth, allowing for variance in size and density profile, though enacting the described detection techniques further afield would be subject to the intent of future space missions.

In summary, we have simulated the behaviour of Earth-traversing UHE neutrinos as affected by the BSM addition of a right-handed neutrino, and demonstrated that the influence of BSM physics on this physical setting (particularly in scenarios that provide a relatively long-lived intermediate particle) might constitute an avenue for probing new physics and the parameter space of BSM theories.

\vspace{10pt}
\section*{acknowledgement}
The authors would like to thank Peter B. Denton for useful suggestions.  We are grateful to Ibrahim Safa and Jeffrey Lazar for their correspondence regarding TauRunner, and acknowledge the great value of the program in the work laid out here. LH acknowledges the support of the Institut Pascal at Université Paris-Saclay during the Paris-Saclay Astroparticle Symposium 2022, with the support of
the IN2P3 master projet UCMN, the P2IO Laboratory of Excellence (program “Investissements d’avenir” ANR-11-IDEX-0003-01 Paris-Saclay and ANR-10-LABX-0038), the P2I axis of the Graduate School Physics of Université Paris-Saclay, as well as IJCLab, CEA, IPhT, APPEC,  and ANR-11-IDEX-0003-01 Paris-Saclay and ANR-10-LABX-0038. LH and RH acknowledge the support of the Science and Technology Facilities Council under Grant ST/T001011/1.

\bibliography{main}

\begin{thebibliography}{62}%
\makeatletter
\providecommand \@ifxundefined [1]{%
 \@ifx{#1\undefined}
}%
\providecommand \@ifnum [1]{%
 \ifnum #1\expandafter \@firstoftwo
 \else \expandafter \@secondoftwo
 \fi
}%
\providecommand \@ifx [1]{%
 \ifx #1\expandafter \@firstoftwo
 \else \expandafter \@secondoftwo
 \fi
}%
\providecommand \natexlab [1]{#1}%
\providecommand \enquote  [1]{``#1''}%
\providecommand \bibnamefont  [1]{#1}%
\providecommand \bibfnamefont [1]{#1}%
\providecommand \citenamefont [1]{#1}%
\providecommand \href@noop [0]{\@secondoftwo}%
\providecommand \href [0]{\begingroup \@sanitize@url \@href}%
\providecommand \@href[1]{\@@startlink{#1}\@@href}%
\providecommand \@@href[1]{\endgroup#1\@@endlink}%
\providecommand \@sanitize@url [0]{\catcode `\\12\catcode `\$12\catcode
  `\&12\catcode `\#12\catcode `\^12\catcode `\_12\catcode `\%12\relax}%
\providecommand \@@startlink[1]{}%
\providecommand \@@endlink[0]{}%
\providecommand \url  [0]{\begingroup\@sanitize@url \@url }%
\providecommand \@url [1]{\endgroup\@href {#1}{\urlprefix }}%
\providecommand \urlprefix  [0]{URL }%
\providecommand \Eprint [0]{\href }%
\providecommand \doibase [0]{http://dx.doi.org/}%
\providecommand \selectlanguage [0]{\@gobble}%
\providecommand \bibinfo  [0]{\@secondoftwo}%
\providecommand \bibfield  [0]{\@secondoftwo}%
\providecommand \translation [1]{[#1]}%
\providecommand \BibitemOpen [0]{}%
\providecommand \bibitemStop [0]{}%
\providecommand \bibitemNoStop [0]{.\EOS\space}%
\providecommand \EOS [0]{\spacefactor3000\relax}%
\providecommand \BibitemShut  [1]{\csname bibitem#1\endcsname}%
\let\auto@bib@innerbib\@empty
\bibitem [{\citenamefont {Venters}\ \emph {et~al.}(2020)\citenamefont
  {Venters}, \citenamefont {Reno}, \citenamefont {Krizmanic}, \citenamefont
  {Anchordoqui}, \citenamefont {Gu\'epin},\ and\ \citenamefont
  {Olinto}}]{Venters:2019xwi}%
  \BibitemOpen
  \bibfield  {author} {\bibinfo {author} {\bibfnamefont {T.~M.}\ \bibnamefont
  {Venters}}, \bibinfo {author} {\bibfnamefont {M.~H.}\ \bibnamefont {Reno}},
  \bibinfo {author} {\bibfnamefont {J.~F.}\ \bibnamefont {Krizmanic}}, \bibinfo
  {author} {\bibfnamefont {L.~A.}\ \bibnamefont {Anchordoqui}}, \bibinfo
  {author} {\bibfnamefont {C.}~\bibnamefont {Gu\'epin}}, \ and\ \bibinfo
  {author} {\bibfnamefont {A.~V.}\ \bibnamefont {Olinto}},\ }\href {\doibase
  10.1103/PhysRevD.102.123013} {\bibfield  {journal} {\bibinfo  {journal}
  {Phys. Rev. D}\ }\textbf {\bibinfo {volume} {102}},\ \bibinfo {pages}
  {123013} (\bibinfo {year} {2020})},\ \Eprint
  {http://arxiv.org/abs/1906.07209} {arXiv:1906.07209 [astro-ph.HE]}
  \BibitemShut {NoStop}%
\bibitem [{\citenamefont {Olinto}\ \emph {et~al.}(2021)\citenamefont {Olinto}
  \emph {et~al.}}]{POEMMA:2020ykm}%
  \BibitemOpen
  \bibfield  {author} {\bibinfo {author} {\bibfnamefont {A.~V.}\ \bibnamefont
  {Olinto}} \emph {et~al.} (\bibinfo {collaboration} {POEMMA}),\ }\href
  {\doibase 10.1088/1475-7516/2021/06/007} {\bibfield  {journal} {\bibinfo
  {journal} {JCAP}\ }\textbf {\bibinfo {volume} {06}},\ \bibinfo {pages} {007}
  (\bibinfo {year} {2021})},\ \Eprint {http://arxiv.org/abs/2012.07945}
  {arXiv:2012.07945 [astro-ph.IM]} \BibitemShut {NoStop}%
\bibitem [{\citenamefont {Lago}(2022)}]{Lago:2021dom}%
  \BibitemOpen
  \bibfield  {author} {\bibinfo {author} {\bibfnamefont {B.~L.}\ \bibnamefont
  {Lago}},\ }\href {\doibase 10.21468/SciPostPhysProc.10.027} {\bibfield
  {journal} {\bibinfo  {journal} {SciPost Phys. Proc.}\ }\textbf {\bibinfo
  {volume} {10}},\ \bibinfo {pages} {027} (\bibinfo {year} {2022})},\ \Eprint
  {http://arxiv.org/abs/2110.14417} {arXiv:2110.14417 [astro-ph.HE]}
  \BibitemShut {NoStop}%
\bibitem [{\citenamefont {Wang}\ \emph {et~al.}(2021)\citenamefont {Wang},
  \citenamefont {Lin}, \citenamefont {Otte}, \citenamefont {Doro},
  \citenamefont {Gazda}, \citenamefont {Taboada}, \citenamefont {Brown},\ and\
  \citenamefont {Bagheri}}]{Wang:2021zkm}%
  \BibitemOpen
  \bibfield  {author} {\bibinfo {author} {\bibfnamefont {A.}~\bibnamefont
  {Wang}}, \bibinfo {author} {\bibfnamefont {C.}~\bibnamefont {Lin}}, \bibinfo
  {author} {\bibfnamefont {N.}~\bibnamefont {Otte}}, \bibinfo {author}
  {\bibfnamefont {M.}~\bibnamefont {Doro}}, \bibinfo {author} {\bibfnamefont
  {E.}~\bibnamefont {Gazda}}, \bibinfo {author} {\bibfnamefont
  {I.}~\bibnamefont {Taboada}}, \bibinfo {author} {\bibfnamefont {A.~M.}\
  \bibnamefont {Brown}}, \ and\ \bibinfo {author} {\bibfnamefont
  {M.}~\bibnamefont {Bagheri}},\ }\href {\doibase 10.22323/1.395.1234}
  {\bibfield  {journal} {\bibinfo  {journal} {PoS}\ }\textbf {\bibinfo {volume}
  {ICRC2021}},\ \bibinfo {pages} {1234} (\bibinfo {year} {2021})},\ \Eprint
  {http://arxiv.org/abs/2108.02751} {arXiv:2108.02751 [astro-ph.IM]}
  \BibitemShut {NoStop}%
\bibitem [{\citenamefont {Zhelnin}\ \emph {et~al.}(2022)\citenamefont
  {Zhelnin}, \citenamefont {Safa}, \citenamefont {Romero-Wolf},\ and\
  \citenamefont {Arg\"uelles}}]{Zhelnin:2022ybr}%
  \BibitemOpen
  \bibfield  {author} {\bibinfo {author} {\bibfnamefont {P.}~\bibnamefont
  {Zhelnin}}, \bibinfo {author} {\bibfnamefont {I.}~\bibnamefont {Safa}},
  \bibinfo {author} {\bibfnamefont {A.}~\bibnamefont {Romero-Wolf}}, \ and\
  \bibinfo {author} {\bibfnamefont {C.~A.}\ \bibnamefont {Arg\"uelles}},\
  }\href {\doibase 10.22323/1.414.0566} {\bibfield  {journal} {\bibinfo
  {journal} {PoS}\ }\textbf {\bibinfo {volume} {ICHEP2022}},\ \bibinfo {pages}
  {566} (\bibinfo {year} {2022})}\BibitemShut {NoStop}%
\bibitem [{\citenamefont {Wang}\ \emph {et~al.}(2022)\citenamefont {Wang} \emph
  {et~al.}}]{TAROGE:2022soh}%
  \BibitemOpen
  \bibfield  {author} {\bibinfo {author} {\bibfnamefont {S.-H.}\ \bibnamefont
  {Wang}} \emph {et~al.} (\bibinfo {collaboration} {TAROGE, Arianna}),\ }\href
  {\doibase 10.1088/1475-7516/2022/11/022} {\bibfield  {journal} {\bibinfo
  {journal} {JCAP}\ }\textbf {\bibinfo {volume} {11}},\ \bibinfo {pages} {022}
  (\bibinfo {year} {2022})},\ \Eprint {http://arxiv.org/abs/2207.10616}
  {arXiv:2207.10616 [astro-ph.HE]} \BibitemShut {NoStop}%
\bibitem [{\citenamefont {Aartsen}\ \emph {et~al.}(2015)\citenamefont {Aartsen}
  \emph {et~al.}}]{IceCube:2015usw}%
  \BibitemOpen
  \bibfield  {author} {\bibinfo {author} {\bibfnamefont {M.~G.}\ \bibnamefont
  {Aartsen}} \emph {et~al.} (\bibinfo {collaboration} {IceCube}),\ }\href
  {\doibase 10.1088/0004-637X/807/1/46} {\bibfield  {journal} {\bibinfo
  {journal} {Astrophys. J.}\ }\textbf {\bibinfo {volume} {807}},\ \bibinfo
  {pages} {46} (\bibinfo {year} {2015})},\ \Eprint
  {http://arxiv.org/abs/1503.00598} {1503.00598 [astro-ph.HE]} \BibitemShut
  {NoStop}%
\bibitem [{\citenamefont {Adrian-Martinez}\ \emph {et~al.}(2015)\citenamefont
  {Adrian-Martinez} \emph {et~al.}}]{ANTARES:2015gxt}%
  \BibitemOpen
  \bibfield  {author} {\bibinfo {author} {\bibfnamefont {S.}~\bibnamefont
  {Adrian-Martinez}} \emph {et~al.} (\bibinfo {collaboration} {ANTARES}),\
  }\href {\doibase 10.1088/1475-7516/2015/12/014} {\bibfield  {journal}
  {\bibinfo  {journal} {JCAP}\ }\textbf {\bibinfo {volume} {12}},\ \bibinfo
  {pages} {014} (\bibinfo {year} {2015})},\ \Eprint
  {http://arxiv.org/abs/1506.07354} {arXiv:1506.07354 [astro-ph.HE]}
  \BibitemShut {NoStop}%
\bibitem [{\citenamefont {Gu\'epin}\ \emph {et~al.}(2022)\citenamefont
  {Gu\'epin}, \citenamefont {Kotera},\ and\ \citenamefont
  {Oikonomou}}]{Guepin:2022qpl}%
  \BibitemOpen
  \bibfield  {author} {\bibinfo {author} {\bibfnamefont {C.}~\bibnamefont
  {Gu\'epin}}, \bibinfo {author} {\bibfnamefont {K.}~\bibnamefont {Kotera}}, \
  and\ \bibinfo {author} {\bibfnamefont {F.}~\bibnamefont {Oikonomou}},\ }\href
  {\doibase 10.1038/s42254-022-00504-9} {\bibfield  {journal} {\bibinfo
  {journal} {Nature Rev. Phys.}\ }\textbf {\bibinfo {volume} {4}},\ \bibinfo
  {pages} {697} (\bibinfo {year} {2022})},\ \Eprint
  {http://arxiv.org/abs/2207.12205} {arXiv:2207.12205 [astro-ph.HE]}
  \BibitemShut {NoStop}%
\bibitem [{\citenamefont {Abarr}\ \emph {et~al.}(2021)\citenamefont {Abarr}
  \emph {et~al.}}]{PUEO:2020bnn}%
  \BibitemOpen
  \bibfield  {author} {\bibinfo {author} {\bibfnamefont {Q.}~\bibnamefont
  {Abarr}} \emph {et~al.} (\bibinfo {collaboration} {PUEO}),\ }\href {\doibase
  10.1088/1748-0221/16/08/P08035} {\bibfield  {journal} {\bibinfo  {journal}
  {JINST}\ }\textbf {\bibinfo {volume} {16}},\ \bibinfo {pages} {P08035}
  (\bibinfo {year} {2021})},\ \Eprint {http://arxiv.org/abs/2010.02892}
  {arXiv:2010.02892 [astro-ph.IM]} \BibitemShut {NoStop}%
\bibitem [{\citenamefont {Romero-Wolf}\ \emph {et~al.}(2019)\citenamefont
  {Romero-Wolf} \emph {et~al.}}]{Romero-Wolf:2018zxt}%
  \BibitemOpen
  \bibfield  {author} {\bibinfo {author} {\bibfnamefont {A.}~\bibnamefont
  {Romero-Wolf}} \emph {et~al.},\ }\href {\doibase 10.1103/PhysRevD.99.063011}
  {\bibfield  {journal} {\bibinfo  {journal} {Phys. Rev. D}\ }\textbf {\bibinfo
  {volume} {99}},\ \bibinfo {pages} {063011} (\bibinfo {year} {2019})},\
  \Eprint {http://arxiv.org/abs/1811.07261} {arXiv:1811.07261 [astro-ph.HE]}
  \BibitemShut {NoStop}%
\bibitem [{\citenamefont {Heurtier}\ \emph
  {et~al.}(2019{\natexlab{a}})\citenamefont {Heurtier}, \citenamefont
  {Mambrini},\ and\ \citenamefont {Pierre}}]{Heurtier:2019git}%
  \BibitemOpen
  \bibfield  {author} {\bibinfo {author} {\bibfnamefont {L.}~\bibnamefont
  {Heurtier}}, \bibinfo {author} {\bibfnamefont {Y.}~\bibnamefont {Mambrini}},
  \ and\ \bibinfo {author} {\bibfnamefont {M.}~\bibnamefont {Pierre}},\ }\href
  {\doibase 10.1103/PhysRevD.99.095014} {\bibfield  {journal} {\bibinfo
  {journal} {Phys. Rev. D}\ }\textbf {\bibinfo {volume} {99}},\ \bibinfo
  {pages} {095014} (\bibinfo {year} {2019}{\natexlab{a}})},\ \Eprint
  {http://arxiv.org/abs/1902.04584} {arXiv:1902.04584 [hep-ph]} \BibitemShut
  {NoStop}%
\bibitem [{\citenamefont {Heurtier}\ \emph
  {et~al.}(2019{\natexlab{b}})\citenamefont {Heurtier}, \citenamefont {Kim},
  \citenamefont {Park},\ and\ \citenamefont {Shin}}]{Heurtier:2019rkz}%
  \BibitemOpen
  \bibfield  {author} {\bibinfo {author} {\bibfnamefont {L.}~\bibnamefont
  {Heurtier}}, \bibinfo {author} {\bibfnamefont {D.}~\bibnamefont {Kim}},
  \bibinfo {author} {\bibfnamefont {J.-C.}\ \bibnamefont {Park}}, \ and\
  \bibinfo {author} {\bibfnamefont {S.}~\bibnamefont {Shin}},\ }\href {\doibase
  10.1103/PhysRevD.100.055004} {\bibfield  {journal} {\bibinfo  {journal}
  {Phys. Rev. D}\ }\textbf {\bibinfo {volume} {100}},\ \bibinfo {pages}
  {055004} (\bibinfo {year} {2019}{\natexlab{b}})},\ \Eprint
  {http://arxiv.org/abs/1905.13223} {arXiv:1905.13223 [hep-ph]} \BibitemShut
  {NoStop}%
\bibitem [{\citenamefont {Anchordoqui}\ \emph {et~al.}(2018)\citenamefont
  {Anchordoqui}, \citenamefont {Barger}, \citenamefont {Learned}, \citenamefont
  {Marfatia},\ and\ \citenamefont {Weiler}}]{Anchordoqui:2018ucj}%
  \BibitemOpen
  \bibfield  {author} {\bibinfo {author} {\bibfnamefont {L.~A.}\ \bibnamefont
  {Anchordoqui}}, \bibinfo {author} {\bibfnamefont {V.}~\bibnamefont {Barger}},
  \bibinfo {author} {\bibfnamefont {J.~G.}\ \bibnamefont {Learned}}, \bibinfo
  {author} {\bibfnamefont {D.}~\bibnamefont {Marfatia}}, \ and\ \bibinfo
  {author} {\bibfnamefont {T.~J.}\ \bibnamefont {Weiler}},\ }\href {\doibase
  10.31526/LHEP.1.2018.03} {\bibfield  {journal} {\bibinfo  {journal} {LHEP}\
  }\textbf {\bibinfo {volume} {1}},\ \bibinfo {pages} {13} (\bibinfo {year}
  {2018})},\ \Eprint {http://arxiv.org/abs/1803.11554} {arXiv:1803.11554
  [hep-ph]} \BibitemShut {NoStop}%
\bibitem [{\citenamefont {Huang}(2018)}]{Huang:2018als}%
  \BibitemOpen
  \bibfield  {author} {\bibinfo {author} {\bibfnamefont {G.-y.}\ \bibnamefont
  {Huang}},\ }\href {\doibase 10.1103/PhysRevD.98.043019} {\bibfield  {journal}
  {\bibinfo  {journal} {Phys. Rev. D}\ }\textbf {\bibinfo {volume} {98}},\
  \bibinfo {pages} {043019} (\bibinfo {year} {2018})},\ \Eprint
  {http://arxiv.org/abs/1804.05362} {arXiv:1804.05362 [hep-ph]} \BibitemShut
  {NoStop}%
\bibitem [{\citenamefont {Chauhan}\ and\ \citenamefont
  {Mohanty}(2019)}]{Chauhan:2018lnq}%
  \BibitemOpen
  \bibfield  {author} {\bibinfo {author} {\bibfnamefont {B.}~\bibnamefont
  {Chauhan}}\ and\ \bibinfo {author} {\bibfnamefont {S.}~\bibnamefont
  {Mohanty}},\ }\href {\doibase 10.1103/PhysRevD.99.095018} {\bibfield
  {journal} {\bibinfo  {journal} {Phys. Rev. D}\ }\textbf {\bibinfo {volume}
  {99}},\ \bibinfo {pages} {095018} (\bibinfo {year} {2019})},\ \Eprint
  {http://arxiv.org/abs/1812.00919} {arXiv:1812.00919 [hep-ph]} \BibitemShut
  {NoStop}%
\bibitem [{\citenamefont {Cherry}\ and\ \citenamefont
  {Shoemaker}(2019)}]{Cherry:2018rxj}%
  \BibitemOpen
  \bibfield  {author} {\bibinfo {author} {\bibfnamefont {J.~F.}\ \bibnamefont
  {Cherry}}\ and\ \bibinfo {author} {\bibfnamefont {I.~M.}\ \bibnamefont
  {Shoemaker}},\ }\href {\doibase 10.1103/PhysRevD.99.063016} {\bibfield
  {journal} {\bibinfo  {journal} {Phys. Rev. D}\ }\textbf {\bibinfo {volume}
  {99}},\ \bibinfo {pages} {063016} (\bibinfo {year} {2019})},\ \Eprint
  {http://arxiv.org/abs/1802.01611} {arXiv:1802.01611 [hep-ph]} \BibitemShut
  {NoStop}%
\bibitem [{\citenamefont {Collins}\ \emph {et~al.}(2019)\citenamefont
  {Collins}, \citenamefont {Bhupal~Dev},\ and\ \citenamefont
  {Sui}}]{Collins:2018jpg}%
  \BibitemOpen
  \bibfield  {author} {\bibinfo {author} {\bibfnamefont {J.~H.}\ \bibnamefont
  {Collins}}, \bibinfo {author} {\bibfnamefont {P.~S.}\ \bibnamefont
  {Bhupal~Dev}}, \ and\ \bibinfo {author} {\bibfnamefont {Y.}~\bibnamefont
  {Sui}},\ }\href {\doibase 10.1103/PhysRevD.99.043009} {\bibfield  {journal}
  {\bibinfo  {journal} {Phys. Rev. D}\ }\textbf {\bibinfo {volume} {99}},\
  \bibinfo {pages} {043009} (\bibinfo {year} {2019})},\ \Eprint
  {http://arxiv.org/abs/1810.08479} {arXiv:1810.08479 [hep-ph]} \BibitemShut
  {NoStop}%
\bibitem [{\citenamefont {Bhupal~Dev}\ \emph {et~al.}(2020)\citenamefont
  {Bhupal~Dev}, \citenamefont {Mohanta}, \citenamefont {Patra},\ and\
  \citenamefont {Sahoo}}]{BhupalDev:2020zcy}%
  \BibitemOpen
  \bibfield  {author} {\bibinfo {author} {\bibfnamefont {P.~S.}\ \bibnamefont
  {Bhupal~Dev}}, \bibinfo {author} {\bibfnamefont {R.}~\bibnamefont {Mohanta}},
  \bibinfo {author} {\bibfnamefont {S.}~\bibnamefont {Patra}}, \ and\ \bibinfo
  {author} {\bibfnamefont {S.}~\bibnamefont {Sahoo}},\ }\href {\doibase
  10.1103/PhysRevD.102.095012} {\bibfield  {journal} {\bibinfo  {journal}
  {Phys. Rev. D}\ }\textbf {\bibinfo {volume} {102}},\ \bibinfo {pages}
  {095012} (\bibinfo {year} {2020})},\ \Eprint
  {http://arxiv.org/abs/2004.09464} {arXiv:2004.09464 [hep-ph]} \BibitemShut
  {NoStop}%
\bibitem [{\citenamefont {Borah}\ \emph {et~al.}(2020)\citenamefont {Borah},
  \citenamefont {Dasgupta}, \citenamefont {Dey},\ and\ \citenamefont
  {Tomar}}]{Borah:2019ciw}%
  \BibitemOpen
  \bibfield  {author} {\bibinfo {author} {\bibfnamefont {D.}~\bibnamefont
  {Borah}}, \bibinfo {author} {\bibfnamefont {A.}~\bibnamefont {Dasgupta}},
  \bibinfo {author} {\bibfnamefont {U.~K.}\ \bibnamefont {Dey}}, \ and\
  \bibinfo {author} {\bibfnamefont {G.}~\bibnamefont {Tomar}},\ }\href
  {\doibase 10.1103/PhysRevD.101.075039} {\bibfield  {journal} {\bibinfo
  {journal} {Phys. Rev. D}\ }\textbf {\bibinfo {volume} {101}},\ \bibinfo
  {pages} {075039} (\bibinfo {year} {2020})},\ \Eprint
  {http://arxiv.org/abs/1907.02740} {arXiv:1907.02740 [hep-ph]} \BibitemShut
  {NoStop}%
\bibitem [{\citenamefont {Cline}\ \emph {et~al.}(2019)\citenamefont {Cline},
  \citenamefont {Gross},\ and\ \citenamefont {Xue}}]{Cline:2019snp}%
  \BibitemOpen
  \bibfield  {author} {\bibinfo {author} {\bibfnamefont {J.~M.}\ \bibnamefont
  {Cline}}, \bibinfo {author} {\bibfnamefont {C.}~\bibnamefont {Gross}}, \ and\
  \bibinfo {author} {\bibfnamefont {W.}~\bibnamefont {Xue}},\ }\href {\doibase
  10.1103/PhysRevD.100.015031} {\bibfield  {journal} {\bibinfo  {journal}
  {Phys. Rev. D}\ }\textbf {\bibinfo {volume} {100}},\ \bibinfo {pages}
  {015031} (\bibinfo {year} {2019})},\ \Eprint
  {http://arxiv.org/abs/1904.13396} {arXiv:1904.13396 [hep-ph]} \BibitemShut
  {NoStop}%
\bibitem [{\citenamefont {Anchordoqui}\ and\ \citenamefont
  {Antoniadis}(2019)}]{Anchordoqui:2018ssd}%
  \BibitemOpen
  \bibfield  {author} {\bibinfo {author} {\bibfnamefont {L.~A.}\ \bibnamefont
  {Anchordoqui}}\ and\ \bibinfo {author} {\bibfnamefont {I.}~\bibnamefont
  {Antoniadis}},\ }\href {\doibase 10.1016/j.physletb.2019.02.003} {\bibfield
  {journal} {\bibinfo  {journal} {Phys. Lett. B}\ }\textbf {\bibinfo {volume}
  {790}},\ \bibinfo {pages} {578} (\bibinfo {year} {2019})},\ \Eprint
  {http://arxiv.org/abs/1812.01520} {arXiv:1812.01520 [hep-ph]} \BibitemShut
  {NoStop}%
\bibitem [{\citenamefont {Esmaili}\ and\ \citenamefont
  {Farzan}(2019)}]{Esmaili:2019pcy}%
  \BibitemOpen
  \bibfield  {author} {\bibinfo {author} {\bibfnamefont {A.}~\bibnamefont
  {Esmaili}}\ and\ \bibinfo {author} {\bibfnamefont {Y.}~\bibnamefont
  {Farzan}},\ }\href {\doibase 10.1088/1475-7516/2019/12/017} {\bibfield
  {journal} {\bibinfo  {journal} {JCAP}\ }\textbf {\bibinfo {volume} {12}},\
  \bibinfo {pages} {017} (\bibinfo {year} {2019})},\ \Eprint
  {http://arxiv.org/abs/1909.07995} {arXiv:1909.07995 [hep-ph]} \BibitemShut
  {NoStop}%
\bibitem [{\citenamefont {Esteban}\ \emph {et~al.}(2020)\citenamefont
  {Esteban}, \citenamefont {Lopez-Pavon}, \citenamefont {Martinez-Soler},\ and\
  \citenamefont {Salvado}}]{Esteban:2019hcm}%
  \BibitemOpen
  \bibfield  {author} {\bibinfo {author} {\bibfnamefont {I.}~\bibnamefont
  {Esteban}}, \bibinfo {author} {\bibfnamefont {J.}~\bibnamefont
  {Lopez-Pavon}}, \bibinfo {author} {\bibfnamefont {I.}~\bibnamefont
  {Martinez-Soler}}, \ and\ \bibinfo {author} {\bibfnamefont {J.}~\bibnamefont
  {Salvado}},\ }\href {\doibase 10.1140/epjc/s10052-020-7816-y} {\bibfield
  {journal} {\bibinfo  {journal} {Eur. Phys. J. C}\ }\textbf {\bibinfo {volume}
  {80}},\ \bibinfo {pages} {259} (\bibinfo {year} {2020})},\ \Eprint
  {http://arxiv.org/abs/1905.10372} {arXiv:1905.10372 [hep-ph]} \BibitemShut
  {NoStop}%
\bibitem [{\citenamefont {Garcia~Soto}\ \emph {et~al.}(2023)\citenamefont
  {Garcia~Soto}, \citenamefont {Garg}, \citenamefont {Reno},\ and\
  \citenamefont {Arg\"uelles}}]{GarciaSoto:2022vlw}%
  \BibitemOpen
  \bibfield  {author} {\bibinfo {author} {\bibfnamefont {A.}~\bibnamefont
  {Garcia~Soto}}, \bibinfo {author} {\bibfnamefont {D.}~\bibnamefont {Garg}},
  \bibinfo {author} {\bibfnamefont {M.~H.}\ \bibnamefont {Reno}}, \ and\
  \bibinfo {author} {\bibfnamefont {C.~A.}\ \bibnamefont {Arg\"uelles}},\
  }\href {\doibase 10.1103/PhysRevD.107.033009} {\bibfield  {journal} {\bibinfo
   {journal} {Phys. Rev. D}\ }\textbf {\bibinfo {volume} {107}},\ \bibinfo
  {pages} {033009} (\bibinfo {year} {2023})},\ \Eprint
  {http://arxiv.org/abs/2209.06282} {arXiv:2209.06282 [hep-ph]} \BibitemShut
  {NoStop}%
\bibitem [{\citenamefont {Denton}\ and\ \citenamefont
  {Kini}(2020)}]{Denton:2020jft}%
  \BibitemOpen
  \bibfield  {author} {\bibinfo {author} {\bibfnamefont {P.~B.}\ \bibnamefont
  {Denton}}\ and\ \bibinfo {author} {\bibfnamefont {Y.}~\bibnamefont {Kini}},\
  }\href {\doibase 10.1103/PhysRevD.102.123019} {\bibfield  {journal} {\bibinfo
   {journal} {Phys. Rev. D}\ }\textbf {\bibinfo {volume} {102}},\ \bibinfo
  {pages} {123019} (\bibinfo {year} {2020})},\ \Eprint
  {http://arxiv.org/abs/2007.10334} {arXiv:2007.10334 [astro-ph.HE]}
  \BibitemShut {NoStop}%
\bibitem [{\citenamefont {Huang}\ \emph {et~al.}(2022)\citenamefont {Huang},
  \citenamefont {Jana}, \citenamefont {Lindner},\ and\ \citenamefont
  {Rodejohann}}]{Huang:2021mki}%
  \BibitemOpen
  \bibfield  {author} {\bibinfo {author} {\bibfnamefont {G.-y.}\ \bibnamefont
  {Huang}}, \bibinfo {author} {\bibfnamefont {S.}~\bibnamefont {Jana}},
  \bibinfo {author} {\bibfnamefont {M.}~\bibnamefont {Lindner}}, \ and\
  \bibinfo {author} {\bibfnamefont {W.}~\bibnamefont {Rodejohann}},\ }\href
  {\doibase 10.1088/1475-7516/2022/02/038} {\bibfield  {journal} {\bibinfo
  {journal} {JCAP}\ }\textbf {\bibinfo {volume} {02}},\ \bibinfo {pages} {038}
  (\bibinfo {year} {2022})},\ \Eprint {http://arxiv.org/abs/2112.09476}
  {arXiv:2112.09476 [hep-ph]} \BibitemShut {NoStop}%
\bibitem [{\citenamefont {Atre}\ \emph {et~al.}(2009)\citenamefont {Atre},
  \citenamefont {Han}, \citenamefont {Pascoli},\ and\ \citenamefont
  {Zhang}}]{Atre:2009rg}%
  \BibitemOpen
  \bibfield  {author} {\bibinfo {author} {\bibfnamefont {A.}~\bibnamefont
  {Atre}}, \bibinfo {author} {\bibfnamefont {T.}~\bibnamefont {Han}}, \bibinfo
  {author} {\bibfnamefont {S.}~\bibnamefont {Pascoli}}, \ and\ \bibinfo
  {author} {\bibfnamefont {B.}~\bibnamefont {Zhang}},\ }\href {\doibase
  10.1088/1126-6708/2009/05/030} {\bibfield  {journal} {\bibinfo  {journal}
  {JHEP}\ }\textbf {\bibinfo {volume} {05}},\ \bibinfo {pages} {030} (\bibinfo
  {year} {2009})},\ \Eprint {http://arxiv.org/abs/0901.3589} {arXiv:0901.3589
  [hep-ph]} \BibitemShut {NoStop}%
\bibitem [{\citenamefont {Formaggio}\ and\ \citenamefont
  {Zeller}(2012)}]{Formaggio:2012}%
  \BibitemOpen
  \bibfield  {author} {\bibinfo {author} {\bibfnamefont {J.~A.}\ \bibnamefont
  {Formaggio}}\ and\ \bibinfo {author} {\bibfnamefont {G.~P.}\ \bibnamefont
  {Zeller}},\ }\href {\doibase 10.1103/RevModPhys.84.1307} {\bibfield
  {journal} {\bibinfo  {journal} {Rev. Mod. Phys.}\ }\textbf {\bibinfo {volume}
  {84}},\ \bibinfo {pages} {1307} (\bibinfo {year} {2012})},\ \Eprint
  {http://arxiv.org/abs/1305.7513} {arXiv:1305.7513 [hep-ex]} \BibitemShut
  {NoStop}%
\bibitem [{\citenamefont {Koehne}\ \emph {et~al.}(2013)\citenamefont {Koehne},
  \citenamefont {Frantzen}, \citenamefont {Schmitz}, \citenamefont {Fuchs},
  \citenamefont {Rhode}, \citenamefont {Chirkin},\ and\ \citenamefont
  {Becker~Tjus}}]{Koehne:2013}%
  \BibitemOpen
  \bibfield  {author} {\bibinfo {author} {\bibfnamefont {J.~H.}\ \bibnamefont
  {Koehne}}, \bibinfo {author} {\bibfnamefont {K.}~\bibnamefont {Frantzen}},
  \bibinfo {author} {\bibfnamefont {M.}~\bibnamefont {Schmitz}}, \bibinfo
  {author} {\bibfnamefont {T.}~\bibnamefont {Fuchs}}, \bibinfo {author}
  {\bibfnamefont {W.}~\bibnamefont {Rhode}}, \bibinfo {author} {\bibfnamefont
  {D.}~\bibnamefont {Chirkin}}, \ and\ \bibinfo {author} {\bibfnamefont
  {J.}~\bibnamefont {Becker~Tjus}},\ }\href {\doibase
  10.1016/j.cpc.2013.04.001} {\bibfield  {journal} {\bibinfo  {journal}
  {Comput. Phys. Commun.}\ }\textbf {\bibinfo {volume} {184}},\ \bibinfo
  {pages} {2070} (\bibinfo {year} {2013})}\BibitemShut {NoStop}%
\bibitem [{\citenamefont {Safa}\ \emph {et~al.}(2022)\citenamefont {Safa},
  \citenamefont {Lazar}, \citenamefont {Pizzuto}, \citenamefont {Vasquez},
  \citenamefont {Arg\"uelles},\ and\ \citenamefont
  {Vandenbroucke}}]{Safa:2022}%
  \BibitemOpen
  \bibfield  {author} {\bibinfo {author} {\bibfnamefont {I.}~\bibnamefont
  {Safa}}, \bibinfo {author} {\bibfnamefont {J.}~\bibnamefont {Lazar}},
  \bibinfo {author} {\bibfnamefont {A.}~\bibnamefont {Pizzuto}}, \bibinfo
  {author} {\bibfnamefont {O.}~\bibnamefont {Vasquez}}, \bibinfo {author}
  {\bibfnamefont {C.~A.}\ \bibnamefont {Arg\"uelles}}, \ and\ \bibinfo {author}
  {\bibfnamefont {J.}~\bibnamefont {Vandenbroucke}},\ }\href {\doibase
  10.1016/j.cpc.2022.108422} {\bibfield  {journal} {\bibinfo  {journal}
  {Comput. Phys. Commun.}\ }\textbf {\bibinfo {volume} {278}},\ \bibinfo
  {pages} {108422} (\bibinfo {year} {2022})},\ \Eprint
  {http://arxiv.org/abs/2110.14662} {arXiv:2110.14662 [hep-ph]} \BibitemShut
  {NoStop}%
\bibitem [{\citenamefont {Safa}\ \emph {et~al.}(2020)\citenamefont {Safa},
  \citenamefont {Pizzuto}, \citenamefont {Arg\"uelles}, \citenamefont {Halzen},
  \citenamefont {Hussain}, \citenamefont {Kheirandish},\ and\ \citenamefont
  {Vandenbroucke}}]{Safa:2020}%
  \BibitemOpen
  \bibfield  {author} {\bibinfo {author} {\bibfnamefont {I.}~\bibnamefont
  {Safa}}, \bibinfo {author} {\bibfnamefont {A.}~\bibnamefont {Pizzuto}},
  \bibinfo {author} {\bibfnamefont {C.~A.}\ \bibnamefont {Arg\"uelles}},
  \bibinfo {author} {\bibfnamefont {F.}~\bibnamefont {Halzen}}, \bibinfo
  {author} {\bibfnamefont {R.}~\bibnamefont {Hussain}}, \bibinfo {author}
  {\bibfnamefont {A.}~\bibnamefont {Kheirandish}}, \ and\ \bibinfo {author}
  {\bibfnamefont {J.}~\bibnamefont {Vandenbroucke}},\ }\href {\doibase
  10.1088/1475-7516/2020/01/012} {\bibfield  {journal} {\bibinfo  {journal}
  {JCAP}\ }\textbf {\bibinfo {volume} {01}},\ \bibinfo {pages} {012} (\bibinfo
  {year} {2020})},\ \Eprint {http://arxiv.org/abs/1909.10487} {arXiv:1909.10487
  [hep-ph]} \BibitemShut {NoStop}%
\bibitem [{\citenamefont {\'Alvarez-Mu\~niz}\ \emph {et~al.}(2020)\citenamefont
  {\'Alvarez-Mu\~niz} \emph {et~al.}}]{GRAND:2018iaj}%
  \BibitemOpen
  \bibfield  {author} {\bibinfo {author} {\bibfnamefont {J.}~\bibnamefont
  {\'Alvarez-Mu\~niz}} \emph {et~al.} (\bibinfo {collaboration} {GRAND}),\
  }\href {\doibase 10.1007/s11433-018-9385-7} {\bibfield  {journal} {\bibinfo
  {journal} {Sci. China Phys. Mech. Astron.}\ }\textbf {\bibinfo {volume}
  {63}},\ \bibinfo {pages} {219501} (\bibinfo {year} {2020})},\ \Eprint
  {http://arxiv.org/abs/1810.09994} {arXiv:1810.09994 [astro-ph.HE]}
  \BibitemShut {NoStop}%
\bibitem [{\citenamefont {Aab}\ \emph {et~al.}(2019)\citenamefont {Aab} \emph
  {et~al.}}]{PierreAuger:2019ens}%
  \BibitemOpen
  \bibfield  {author} {\bibinfo {author} {\bibfnamefont {A.}~\bibnamefont
  {Aab}} \emph {et~al.} (\bibinfo {collaboration} {Pierre Auger}),\ }\href
  {\doibase 10.1088/1475-7516/2019/10/022} {\bibfield  {journal} {\bibinfo
  {journal} {JCAP}\ }\textbf {\bibinfo {volume} {10}},\ \bibinfo {pages} {022}
  (\bibinfo {year} {2019})},\ \Eprint {http://arxiv.org/abs/1906.07422}
  {arXiv:1906.07422 [astro-ph.HE]} \BibitemShut {NoStop}%
\bibitem [{\citenamefont {Dienes}\ and\ \citenamefont
  {Thomas}(2012)}]{Dienes:2011ja}%
  \BibitemOpen
  \bibfield  {author} {\bibinfo {author} {\bibfnamefont {K.~R.}\ \bibnamefont
  {Dienes}}\ and\ \bibinfo {author} {\bibfnamefont {B.}~\bibnamefont
  {Thomas}},\ }\href {\doibase 10.1103/PhysRevD.85.083523} {\bibfield
  {journal} {\bibinfo  {journal} {Phys. Rev. D}\ }\textbf {\bibinfo {volume}
  {85}},\ \bibinfo {pages} {083523} (\bibinfo {year} {2012})},\ \Eprint
  {http://arxiv.org/abs/1106.4546} {arXiv:1106.4546 [hep-ph]} \BibitemShut
  {NoStop}%
\bibitem [{\citenamefont {Dienes}\ \emph {et~al.}(2022)\citenamefont {Dienes},
  \citenamefont {Heurtier}, \citenamefont {Huang}, \citenamefont {Kim},
  \citenamefont {Tait},\ and\ \citenamefont {Thomas}}]{Dienes:2021woi}%
  \BibitemOpen
  \bibfield  {author} {\bibinfo {author} {\bibfnamefont {K.~R.}\ \bibnamefont
  {Dienes}}, \bibinfo {author} {\bibfnamefont {L.}~\bibnamefont {Heurtier}},
  \bibinfo {author} {\bibfnamefont {F.}~\bibnamefont {Huang}}, \bibinfo
  {author} {\bibfnamefont {D.}~\bibnamefont {Kim}}, \bibinfo {author}
  {\bibfnamefont {T.~M.~P.}\ \bibnamefont {Tait}}, \ and\ \bibinfo {author}
  {\bibfnamefont {B.}~\bibnamefont {Thomas}},\ }\href {\doibase
  10.1103/PhysRevD.105.023530} {\bibfield  {journal} {\bibinfo  {journal}
  {Phys. Rev. D}\ }\textbf {\bibinfo {volume} {105}},\ \bibinfo {pages}
  {023530} (\bibinfo {year} {2022})},\ \Eprint
  {http://arxiv.org/abs/2111.04753} {arXiv:2111.04753 [astro-ph.CO]}
  \BibitemShut {NoStop}%
\bibitem [{\citenamefont {Pilaftsis}(1999)}]{Pilaftsis:1999jk}%
  \BibitemOpen
  \bibfield  {author} {\bibinfo {author} {\bibfnamefont {A.}~\bibnamefont
  {Pilaftsis}},\ }\href {\doibase 10.1103/PhysRevD.60.105023} {\bibfield
  {journal} {\bibinfo  {journal} {Phys. Rev. D}\ }\textbf {\bibinfo {volume}
  {60}},\ \bibinfo {pages} {105023} (\bibinfo {year} {1999})},\ \Eprint
  {http://arxiv.org/abs/hep-ph/9906265} {arXiv:hep-ph/9906265} \BibitemShut
  {NoStop}%
\bibitem [{\citenamefont {Dienes}\ \emph {et~al.}(1999)\citenamefont {Dienes},
  \citenamefont {Dudas},\ and\ \citenamefont {Gherghetta}}]{Dienes:1998sb}%
  \BibitemOpen
  \bibfield  {author} {\bibinfo {author} {\bibfnamefont {K.~R.}\ \bibnamefont
  {Dienes}}, \bibinfo {author} {\bibfnamefont {E.}~\bibnamefont {Dudas}}, \
  and\ \bibinfo {author} {\bibfnamefont {T.}~\bibnamefont {Gherghetta}},\
  }\href {\doibase 10.1016/S0550-3213(99)00377-6} {\bibfield  {journal}
  {\bibinfo  {journal} {Nucl. Phys. B}\ }\textbf {\bibinfo {volume} {557}},\
  \bibinfo {pages} {25} (\bibinfo {year} {1999})},\ \Eprint
  {http://arxiv.org/abs/hep-ph/9811428} {arXiv:hep-ph/9811428} \BibitemShut
  {NoStop}%
\bibitem [{\citenamefont {Olinto}\ \emph {et~al.}(2018)\citenamefont {Olinto}
  \emph {et~al.}}]{Olinto:2017xbi}%
  \BibitemOpen
  \bibfield  {author} {\bibinfo {author} {\bibfnamefont {A.~V.}\ \bibnamefont
  {Olinto}} \emph {et~al.},\ }\href {\doibase 10.22323/1.301.0542} {\bibfield
  {journal} {\bibinfo  {journal} {PoS}\ }\textbf {\bibinfo {volume}
  {ICRC2017}},\ \bibinfo {pages} {542} (\bibinfo {year} {2018})},\ \Eprint
  {http://arxiv.org/abs/1708.07599} {arXiv:1708.07599 [astro-ph.IM]}
  \BibitemShut {NoStop}%
\bibitem [{\citenamefont {Reno}\ \emph {et~al.}(2019)\citenamefont {Reno},
  \citenamefont {Krizmanic},\ and\ \citenamefont {Venters}}]{Reno:2019jtr}%
  \BibitemOpen
  \bibfield  {author} {\bibinfo {author} {\bibfnamefont {M.~H.}\ \bibnamefont
  {Reno}}, \bibinfo {author} {\bibfnamefont {J.~F.}\ \bibnamefont {Krizmanic}},
  \ and\ \bibinfo {author} {\bibfnamefont {T.~M.}\ \bibnamefont {Venters}},\
  }\href {\doibase 10.1103/PhysRevD.100.063010} {\bibfield  {journal} {\bibinfo
   {journal} {Phys. Rev. D}\ }\textbf {\bibinfo {volume} {100}},\ \bibinfo
  {pages} {063010} (\bibinfo {year} {2019})},\ \Eprint
  {http://arxiv.org/abs/1902.11287} {arXiv:1902.11287 [astro-ph.HE]}
  \BibitemShut {NoStop}%
\bibitem [{\citenamefont {Cummings}\ \emph {et~al.}(2021)\citenamefont
  {Cummings}, \citenamefont {Aloisio},\ and\ \citenamefont
  {Krizmanic}}]{Cummings:2020ycz}%
  \BibitemOpen
  \bibfield  {author} {\bibinfo {author} {\bibfnamefont {A.~L.}\ \bibnamefont
  {Cummings}}, \bibinfo {author} {\bibfnamefont {R.}~\bibnamefont {Aloisio}}, \
  and\ \bibinfo {author} {\bibfnamefont {J.~F.}\ \bibnamefont {Krizmanic}},\
  }\href {\doibase 10.1103/PhysRevD.103.043017} {\bibfield  {journal} {\bibinfo
   {journal} {Phys. Rev. D}\ }\textbf {\bibinfo {volume} {103}},\ \bibinfo
  {pages} {043017} (\bibinfo {year} {2021})},\ \Eprint
  {http://arxiv.org/abs/2011.09869} {arXiv:2011.09869 [astro-ph.HE]}
  \BibitemShut {NoStop}%
\bibitem [{\citenamefont {Krizmanic}(2021)}]{Krizmanic:2020shl}%
  \BibitemOpen
  \bibfield  {author} {\bibinfo {author} {\bibfnamefont {J.~F.}\ \bibnamefont
  {Krizmanic}} (\bibinfo {collaboration} {POEMMA}),\ }\href {\doibase
  10.1016/j.nima.2020.164614} {\bibfield  {journal} {\bibinfo  {journal} {Nucl.
  Instrum. Meth. A}\ }\textbf {\bibinfo {volume} {985}},\ \bibinfo {pages}
  {164614} (\bibinfo {year} {2021})},\ \Eprint
  {http://arxiv.org/abs/2008.04984} {arXiv:2008.04984 [astro-ph.IM]}
  \BibitemShut {NoStop}%
\bibitem [{\citenamefont {Kakimoto}\ \emph {et~al.}(1996)\citenamefont
  {Kakimoto}, \citenamefont {Loh}, \citenamefont {Nagano}, \citenamefont
  {Okuno}, \citenamefont {Teshima},\ and\ \citenamefont
  {Ueno}}]{Kakimoto:1995pr}%
  \BibitemOpen
  \bibfield  {author} {\bibinfo {author} {\bibfnamefont {F.}~\bibnamefont
  {Kakimoto}}, \bibinfo {author} {\bibfnamefont {E.~C.}\ \bibnamefont {Loh}},
  \bibinfo {author} {\bibfnamefont {M.}~\bibnamefont {Nagano}}, \bibinfo
  {author} {\bibfnamefont {H.}~\bibnamefont {Okuno}}, \bibinfo {author}
  {\bibfnamefont {M.}~\bibnamefont {Teshima}}, \ and\ \bibinfo {author}
  {\bibfnamefont {S.}~\bibnamefont {Ueno}},\ }\href {\doibase
  10.1016/0168-9002(95)01423-3} {\bibfield  {journal} {\bibinfo  {journal}
  {Nucl. Instrum. Meth. A}\ }\textbf {\bibinfo {volume} {372}},\ \bibinfo
  {pages} {527} (\bibinfo {year} {1996})}\BibitemShut {NoStop}%
\bibitem [{\citenamefont {Rudolph}\ \emph {et~al.}(2023)\citenamefont
  {Rudolph}, \citenamefont {Petropoulou}, \citenamefont {Winter},\ and\
  \citenamefont {Bo\v{s}njak}}]{Rudolph:2022dky}%
  \BibitemOpen
  \bibfield  {author} {\bibinfo {author} {\bibfnamefont {A.}~\bibnamefont
  {Rudolph}}, \bibinfo {author} {\bibfnamefont {M.}~\bibnamefont
  {Petropoulou}}, \bibinfo {author} {\bibfnamefont {W.}~\bibnamefont {Winter}},
  \ and\ \bibinfo {author} {\bibfnamefont {v.}~\bibnamefont {Bo\v{s}njak}},\
  }\href {\doibase 10.3847/2041-8213/acb6d7} {\bibfield  {journal} {\bibinfo
  {journal} {Astrophys. J. Lett.}\ }\textbf {\bibinfo {volume} {944}},\
  \bibinfo {pages} {L34} (\bibinfo {year} {2023})},\ \Eprint
  {http://arxiv.org/abs/2212.00766} {arXiv:2212.00766 [astro-ph.HE]}
  \BibitemShut {NoStop}%
\bibitem [{\citenamefont {Gorham}\ \emph {et~al.}(2018)\citenamefont {Gorham}
  \emph {et~al.}}]{ANITA:2018vwl}%
  \BibitemOpen
  \bibfield  {author} {\bibinfo {author} {\bibfnamefont {P.~W.}\ \bibnamefont
  {Gorham}} \emph {et~al.} (\bibinfo {collaboration} {ANITA}),\ }\href
  {\doibase 10.1103/PhysRevD.98.022001} {\bibfield  {journal} {\bibinfo
  {journal} {Phys. Rev. D}\ }\textbf {\bibinfo {volume} {98}},\ \bibinfo
  {pages} {022001} (\bibinfo {year} {2018})},\ \Eprint
  {http://arxiv.org/abs/1803.02719} {arXiv:1803.02719 [astro-ph.HE]}
  \BibitemShut {NoStop}%
\bibitem [{\citenamefont {Aartsen}\ \emph {et~al.}(2018)\citenamefont {Aartsen}
  \emph {et~al.}}]{IceCube:2018fhm}%
  \BibitemOpen
  \bibfield  {author} {\bibinfo {author} {\bibfnamefont {M.~G.}\ \bibnamefont
  {Aartsen}} \emph {et~al.} (\bibinfo {collaboration} {IceCube}),\ }\href
  {\doibase 10.1103/PhysRevD.98.062003} {\bibfield  {journal} {\bibinfo
  {journal} {Phys. Rev. D}\ }\textbf {\bibinfo {volume} {98}},\ \bibinfo
  {pages} {062003} (\bibinfo {year} {2018})},\ \Eprint
  {http://arxiv.org/abs/1807.01820} {arXiv:1807.01820 [astro-ph.HE]}
  \BibitemShut {NoStop}%
\bibitem [{\citenamefont {Abbott}\ \emph {et~al.}(2017)\citenamefont {Abbott}
  \emph {et~al.}}]{LIGOScientific:2017vwq}%
  \BibitemOpen
  \bibfield  {author} {\bibinfo {author} {\bibfnamefont {B.~P.}\ \bibnamefont
  {Abbott}} \emph {et~al.} (\bibinfo {collaboration} {LIGO Scientific,
  Virgo}),\ }\href {\doibase 10.1103/PhysRevLett.119.161101} {\bibfield
  {journal} {\bibinfo  {journal} {Phys. Rev. Lett.}\ }\textbf {\bibinfo
  {volume} {119}},\ \bibinfo {pages} {161101} (\bibinfo {year} {2017})},\
  \Eprint {http://arxiv.org/abs/1710.05832} {arXiv:1710.05832 [gr-qc]}
  \BibitemShut {NoStop}%
\bibitem [{\citenamefont {Waxman}(1995)}]{Waxman:1995vg}%
  \BibitemOpen
  \bibfield  {author} {\bibinfo {author} {\bibfnamefont {E.}~\bibnamefont
  {Waxman}},\ }\href {\doibase 10.1103/PhysRevLett.75.386} {\bibfield
  {journal} {\bibinfo  {journal} {Phys. Rev. Lett.}\ }\textbf {\bibinfo
  {volume} {75}},\ \bibinfo {pages} {386} (\bibinfo {year} {1995})},\ \Eprint
  {http://arxiv.org/abs/astro-ph/9505082} {arXiv:astro-ph/9505082} \BibitemShut
  {NoStop}%
\bibitem [{\citenamefont {Vietri}(1995)}]{Vietri:1995hs}%
  \BibitemOpen
  \bibfield  {author} {\bibinfo {author} {\bibfnamefont {M.}~\bibnamefont
  {Vietri}},\ }\href {\doibase 10.1086/176448} {\bibfield  {journal} {\bibinfo
  {journal} {Astrophys. J.}\ }\textbf {\bibinfo {volume} {453}},\ \bibinfo
  {pages} {883} (\bibinfo {year} {1995})},\ \Eprint
  {http://arxiv.org/abs/astro-ph/9506081} {arXiv:astro-ph/9506081} \BibitemShut
  {NoStop}%
\bibitem [{\citenamefont {{de Ugarte Postigo}}\ \emph
  {et~al.}(2022)\citenamefont {{de Ugarte Postigo}}, \citenamefont {{Izzo}},
  \citenamefont {{Pugliese}}, \citenamefont {{Xu}}, \citenamefont
  {{Schneider}}, \citenamefont {{Fynbo}}, \citenamefont {{Tanvir}},
  \citenamefont {{Malesani}}, \citenamefont {{Saccardi}}, \citenamefont
  {{Kann}}, \citenamefont {{Wiersema}}, \citenamefont {{Gompertz}},
  \citenamefont {{Thoene}}, \citenamefont {{Levan}},\ and\ \citenamefont
  {{Stargate Collaboration}}}]{2022GCN.32648....1D}%
  \BibitemOpen
  \bibfield  {author} {\bibinfo {author} {\bibfnamefont {A.}~\bibnamefont {{de
  Ugarte Postigo}}}, \bibinfo {author} {\bibfnamefont {L.}~\bibnamefont
  {{Izzo}}}, \bibinfo {author} {\bibfnamefont {G.}~\bibnamefont {{Pugliese}}},
  \bibinfo {author} {\bibfnamefont {D.}~\bibnamefont {{Xu}}}, \bibinfo {author}
  {\bibfnamefont {B.}~\bibnamefont {{Schneider}}}, \bibinfo {author}
  {\bibfnamefont {J.~P.~U.}\ \bibnamefont {{Fynbo}}}, \bibinfo {author}
  {\bibfnamefont {N.~R.}\ \bibnamefont {{Tanvir}}}, \bibinfo {author}
  {\bibfnamefont {D.~B.}\ \bibnamefont {{Malesani}}}, \bibinfo {author}
  {\bibfnamefont {A.}~\bibnamefont {{Saccardi}}}, \bibinfo {author}
  {\bibfnamefont {D.~A.}\ \bibnamefont {{Kann}}}, \bibinfo {author}
  {\bibfnamefont {K.}~\bibnamefont {{Wiersema}}}, \bibinfo {author}
  {\bibfnamefont {B.~P.}\ \bibnamefont {{Gompertz}}}, \bibinfo {author}
  {\bibfnamefont {C.~C.}\ \bibnamefont {{Thoene}}}, \bibinfo {author}
  {\bibfnamefont {A.~J.}\ \bibnamefont {{Levan}}}, \ and\ \bibinfo {author}
  {\bibnamefont {{Stargate Collaboration}}},\ }\href@noop {} {\bibfield
  {journal} {\bibinfo  {journal} {GRB Coordinates Network}\ }\textbf {\bibinfo
  {volume} {32648}},\ \bibinfo {pages} {1} (\bibinfo {year}
  {2022})}\BibitemShut {NoStop}%
\bibitem [{\citenamefont {{Veres}}\ \emph {et~al.}(2022)\citenamefont
  {{Veres}}, \citenamefont {{Burns}}, \citenamefont {{Bissaldi}}, \citenamefont
  {{Lesage}}, \citenamefont {{Roberts}},\ and\ \citenamefont {{Fermi GBM
  Team}}}]{2022GCN.32636....1V}%
  \BibitemOpen
  \bibfield  {author} {\bibinfo {author} {\bibfnamefont {P.}~\bibnamefont
  {{Veres}}}, \bibinfo {author} {\bibfnamefont {E.}~\bibnamefont {{Burns}}},
  \bibinfo {author} {\bibfnamefont {E.}~\bibnamefont {{Bissaldi}}}, \bibinfo
  {author} {\bibfnamefont {S.}~\bibnamefont {{Lesage}}}, \bibinfo {author}
  {\bibfnamefont {O.}~\bibnamefont {{Roberts}}}, \ and\ \bibinfo {author}
  {\bibnamefont {{Fermi GBM Team}}},\ }\href@noop {} {\bibfield  {journal}
  {\bibinfo  {journal} {GRB Coordinates Network}\ }\textbf {\bibinfo {volume}
  {32636}},\ \bibinfo {pages} {1} (\bibinfo {year} {2022})}\BibitemShut
  {NoStop}%
\bibitem [{\citenamefont {{Dichiara}}\ \emph {et~al.}(2022)\citenamefont
  {{Dichiara}}, \citenamefont {{Gropp}}, \citenamefont {{Kennea}},
  \citenamefont {{Kuin}}, \citenamefont {{Lien}}, \citenamefont {{Marshall}},
  \citenamefont {{Tohuvavohu}},\ and\ \citenamefont
  {{Williams}}}]{2022ATel15650....1D}%
  \BibitemOpen
  \bibfield  {author} {\bibinfo {author} {\bibfnamefont {S.}~\bibnamefont
  {{Dichiara}}}, \bibinfo {author} {\bibfnamefont {J.~D.}\ \bibnamefont
  {{Gropp}}}, \bibinfo {author} {\bibfnamefont {J.~A.}\ \bibnamefont
  {{Kennea}}}, \bibinfo {author} {\bibfnamefont {N.~P.~M.}\ \bibnamefont
  {{Kuin}}}, \bibinfo {author} {\bibfnamefont {A.~Y.}\ \bibnamefont {{Lien}}},
  \bibinfo {author} {\bibfnamefont {F.~E.}\ \bibnamefont {{Marshall}}},
  \bibinfo {author} {\bibfnamefont {A.}~\bibnamefont {{Tohuvavohu}}}, \ and\
  \bibinfo {author} {\bibfnamefont {M.~A.}\ \bibnamefont {{Williams}}},\
  }\href@noop {} {\bibfield  {journal} {\bibinfo  {journal} {The Astronomer's
  Telegram}\ }\textbf {\bibinfo {volume} {15650}},\ \bibinfo {pages} {1}
  (\bibinfo {year} {2022})}\BibitemShut {NoStop}%
\bibitem [{\citenamefont {Pillera}\ \emph {et~al.}(2022)\citenamefont
  {Pillera}, \citenamefont {Bissaldi}, \citenamefont {Omodei}, \citenamefont
  {La~Mura}, \citenamefont {Longo}, \citenamefont {team} \emph
  {et~al.}}]{pillera2022grb}%
  \BibitemOpen
  \bibfield  {author} {\bibinfo {author} {\bibfnamefont {R.}~\bibnamefont
  {Pillera}}, \bibinfo {author} {\bibfnamefont {E.}~\bibnamefont {Bissaldi}},
  \bibinfo {author} {\bibfnamefont {N.}~\bibnamefont {Omodei}}, \bibinfo
  {author} {\bibfnamefont {G.}~\bibnamefont {La~Mura}}, \bibinfo {author}
  {\bibfnamefont {F.}~\bibnamefont {Longo}}, \bibinfo {author} {\bibfnamefont
  {F.-L.}\ \bibnamefont {team}},  \emph {et~al.},\ }\href@noop {} {\bibfield
  {journal} {\bibinfo  {journal} {GRB Coordinates Network}\ }\textbf {\bibinfo
  {volume} {32658}},\ \bibinfo {pages} {1} (\bibinfo {year}
  {2022})}\BibitemShut {NoStop}%
\bibitem [{\citenamefont {{Huang}}\ \emph {et~al.}(2022)\citenamefont
  {{Huang}}, \citenamefont {{Hu}}, \citenamefont {{Chen}}, \citenamefont
  {{Zha}}, \citenamefont {{Liu}}, \citenamefont {{Yao}}, \citenamefont
  {{Cao}},\ and\ \citenamefont {{Experiment}}}]{2022GCN.32677....1H}%
  \BibitemOpen
  \bibfield  {author} {\bibinfo {author} {\bibfnamefont {Y.}~\bibnamefont
  {{Huang}}}, \bibinfo {author} {\bibfnamefont {S.}~\bibnamefont {{Hu}}},
  \bibinfo {author} {\bibfnamefont {S.}~\bibnamefont {{Chen}}}, \bibinfo
  {author} {\bibfnamefont {M.}~\bibnamefont {{Zha}}}, \bibinfo {author}
  {\bibfnamefont {C.}~\bibnamefont {{Liu}}}, \bibinfo {author} {\bibfnamefont
  {Z.}~\bibnamefont {{Yao}}}, \bibinfo {author} {\bibfnamefont
  {Z.}~\bibnamefont {{Cao}}}, \ and\ \bibinfo {author} {\bibfnamefont {T.~L.}\
  \bibnamefont {{Experiment}}},\ }\href@noop {} {\bibfield  {journal} {\bibinfo
   {journal} {GRB Coordinates Network}\ }\textbf {\bibinfo {volume} {32677}},\
  \bibinfo {pages} {1} (\bibinfo {year} {2022})}\BibitemShut {NoStop}%
\bibitem [{\citenamefont {Alves~Batista}(2022)}]{AlvesBatista:2022kpg}%
  \BibitemOpen
  \bibfield  {author} {\bibinfo {author} {\bibfnamefont {R.}~\bibnamefont
  {Alves~Batista}},\ }\href@noop {} {\  (\bibinfo {year} {2022})},\ \Eprint
  {http://arxiv.org/abs/2210.12855} {arXiv:2210.12855 [astro-ph.HE]}
  \BibitemShut {NoStop}%
\bibitem [{\citenamefont {{IceCube
  Collaboration}}(2022)}]{2022GCN.32665....1I}%
  \BibitemOpen
  \bibfield  {author} {\bibinfo {author} {\bibnamefont {{IceCube
  Collaboration}}},\ }\href@noop {} {\bibfield  {journal} {\bibinfo  {journal}
  {GRB Coordinates Network}\ }\textbf {\bibinfo {volume} {32665}},\ \bibinfo
  {pages} {1} (\bibinfo {year} {2022})}\BibitemShut {NoStop}%
\bibitem [{\citenamefont {Bergsma}\ \emph {et~al.}(1985)\citenamefont {Bergsma}
  \emph {et~al.}}]{CHARM:1985anb}%
  \BibitemOpen
  \bibfield  {author} {\bibinfo {author} {\bibfnamefont {F.}~\bibnamefont
  {Bergsma}} \emph {et~al.} (\bibinfo {collaboration} {CHARM}),\ }\href
  {\doibase 10.1016/0370-2693(85)90400-9} {\bibfield  {journal} {\bibinfo
  {journal} {Phys. Lett. B}\ }\textbf {\bibinfo {volume} {157}},\ \bibinfo
  {pages} {458} (\bibinfo {year} {1985})}\BibitemShut {NoStop}%
\bibitem [{\citenamefont {Abreu}\ \emph {et~al.}(1997)\citenamefont {Abreu}
  \emph {et~al.}}]{DELPHI:1996qcc}%
  \BibitemOpen
  \bibfield  {author} {\bibinfo {author} {\bibfnamefont {P.}~\bibnamefont
  {Abreu}} \emph {et~al.} (\bibinfo {collaboration} {DELPHI}),\ }\href
  {\doibase 10.1007/s002880050370} {\bibfield  {journal} {\bibinfo  {journal}
  {Z. Phys. C}\ }\textbf {\bibinfo {volume} {74}},\ \bibinfo {pages} {57}
  (\bibinfo {year} {1997})},\ \bibinfo {note} {[Erratum: Z.Phys.C 75, 580
  (1997)]}\BibitemShut {NoStop}%
\bibitem [{\citenamefont {Ahdida}\ \emph {et~al.}(2019)\citenamefont {Ahdida}
  \emph {et~al.}}]{SHiP:2018xqw}%
  \BibitemOpen
  \bibfield  {author} {\bibinfo {author} {\bibfnamefont {C.}~\bibnamefont
  {Ahdida}} \emph {et~al.} (\bibinfo {collaboration} {SHiP}),\ }\href {\doibase
  10.1007/JHEP04(2019)077} {\bibfield  {journal} {\bibinfo  {journal} {JHEP}\
  }\textbf {\bibinfo {volume} {04}},\ \bibinfo {pages} {077} (\bibinfo {year}
  {2019})},\ \Eprint {http://arxiv.org/abs/1811.00930} {arXiv:1811.00930
  [hep-ph]} \BibitemShut {NoStop}%
\bibitem [{\citenamefont {Cheung}\ \emph {et~al.}(2020)\citenamefont {Cheung},
  \citenamefont {Chung}, \citenamefont {Ishida},\ and\ \citenamefont
  {Lu}}]{Cheung:2020buy}%
  \BibitemOpen
  \bibfield  {author} {\bibinfo {author} {\bibfnamefont {K.}~\bibnamefont
  {Cheung}}, \bibinfo {author} {\bibfnamefont {Y.-L.}\ \bibnamefont {Chung}},
  \bibinfo {author} {\bibfnamefont {H.}~\bibnamefont {Ishida}}, \ and\ \bibinfo
  {author} {\bibfnamefont {C.-T.}\ \bibnamefont {Lu}},\ }\href {\doibase
  10.1103/PhysRevD.102.075038} {\bibfield  {journal} {\bibinfo  {journal}
  {Phys. Rev. D}\ }\textbf {\bibinfo {volume} {102}},\ \bibinfo {pages}
  {075038} (\bibinfo {year} {2020})},\ \Eprint
  {http://arxiv.org/abs/2004.11537} {arXiv:2004.11537 [hep-ph]} \BibitemShut
  {NoStop}%
\bibitem [{\citenamefont {Fischer}\ \emph {et~al.}(2023)\citenamefont
  {Fischer}, \citenamefont {Pattnaik},\ and\ \citenamefont
  {Zurita}}]{Fischer:2023bfn}%
  \BibitemOpen
  \bibfield  {author} {\bibinfo {author} {\bibfnamefont {O.}~\bibnamefont
  {Fischer}}, \bibinfo {author} {\bibfnamefont {B.}~\bibnamefont {Pattnaik}}, \
  and\ \bibinfo {author} {\bibfnamefont {J.}~\bibnamefont {Zurita}},\
  }\href@noop {} {\  (\bibinfo {year} {2023})},\ \Eprint
  {http://arxiv.org/abs/2301.07120} {arXiv:2301.07120 [hep-ph]} \BibitemShut
  {NoStop}%
\bibitem [{\citenamefont {Soto}\ \emph {et~al.}(2022)\citenamefont {Soto},
  \citenamefont {Zhelnin}, \citenamefont {Safa},\ and\ \citenamefont
  {Arg\"uelles}}]{Soto:2021vdc}%
  \BibitemOpen
  \bibfield  {author} {\bibinfo {author} {\bibfnamefont {A.~G.}\ \bibnamefont
  {Soto}}, \bibinfo {author} {\bibfnamefont {P.}~\bibnamefont {Zhelnin}},
  \bibinfo {author} {\bibfnamefont {I.}~\bibnamefont {Safa}}, \ and\ \bibinfo
  {author} {\bibfnamefont {C.~A.}\ \bibnamefont {Arg\"uelles}},\ }\href
  {\doibase 10.1103/PhysRevLett.128.171101} {\bibfield  {journal} {\bibinfo
  {journal} {Phys. Rev. Lett.}\ }\textbf {\bibinfo {volume} {128}},\ \bibinfo
  {pages} {171101} (\bibinfo {year} {2022})},\ \Eprint
  {http://arxiv.org/abs/2112.06937} {arXiv:2112.06937 [hep-ph]} \BibitemShut
  {NoStop}%
\end{thebibliography}%

\end{document}